\documentstyle[11pt]{article}

\catcode`@=11 
\def\lsim{\mathrel{\mathpalette\@versim<}}
\def\gsim{\mathrel{\mathpalette\@versim>}}
\def\@versim#1#2{\lower0.2ex\vbox{\baselineskip\z@skip
       \lineskip\z@skip\lineskiplimit\z@\ialign
       {$\m@th#1\hfil##\hfil$\crcr#2\crcr\sim\crcr}}}
\def\noin{\noindent}
\def\non{\nonumber}
\def\beq{\begin{equation}}
\def\eeq{\end{equation}}
\def\ben{\begin{eqnarray}}
\def\een{\end{eqnarray}}
\def\lab{\label}
\def\bib{\bibitem}

\def\etal{{\it et al}}
\def\ra{\rightarrow}

\def\bm{\boldmath}

\def\Enu{E_{\nu}}

\def\q3{{\boldmath $q$}}
\def\C12{^{12}\rm{C}}
\def\N12{^{12}\rm{N}}
\def\B12{^{12}\rm{B}}
\def\C13{^{13}\rm{C}}
\def\N13{^{13}\rm{N}}
\def\O16{^{16}\rm{O}}
\def\Cl37{^{37}\rm{Cl}}
\def\A37{^{37}\rm{Ar}}

\begin{document}
\begin{titlepage}
\noin
USC(NT)-93-6\\
Submitted to {\it Internat'l J. Mod. Phys. E}
\begin{center}
\vspace{0.5cm}

{\large\bf Neutrino-Nucleus Reactions}\\
\vspace{1.0cm}
{\large Kuniharu
Kubodera}\footnote{Work supported in part by
National Science Foundation Grant No. PHYS-9006844.}\\
{\large\it Department of Physics and Astronomy}\\
{\large\it University of South Carolina}\\
{\large\it Columbia, SC 29208, U.S.A.}\\
\vspace{0.5cm}
{\large and}\\
\vspace{0.5cm}
{\large Satoshi Nozawa}\\
{\large\it Department of Physics}\\
{\large\it Queen's University}\\
{\large\it Kingston,Canada, K4L 3N6}
\end{center}
\vspace{0.7in}
\begin{quotation}
\centerline{\large \bf ABSTRACT}

\noin
The current status is reviewed of
theoretical treatments of neutrino-nucleus reactions
that are relevant to the detection of astrophysical neutrinos.
Various nuclear physics aspects involved
in the evaluation of
the neutrino-nucleus reaction cross sections
are critically surveyed.
\end{quotation}
\end{titlepage}

\noin
{\bf Chapter 1\hspace{1cm}Introduction}

Although many topics can come
under the title: ``Neutrino-Nucleus Reactions",
we shall be concerned here with
neutrino-nucleus reactions that pertain
to the terrestrial observation of astrophysical neutrinos.
The neutrinos play fundamental roles in
various astrophysical phenomena.
The wealth of these phenomena
is eruditely described in, {\it e.g.},
Bahcall \cite{bah89} and Fukugita \cite{fuk87}.
Obviously, terrestrial experiments to detect these neutrinos
are highly valuable sources of astrophysical information,
and observational neutrino astrophysics
is an extremely rapidly expanding research field
\cite{kos92}.
It is useful to distinguish three different energy
regimes of the astrophysical neutrinos.
(i) The low energy region ($E_{\nu} \lsim 20 \rm{MeV}$),
which includes the solar neutrinos
and the lower energy part of the supernova neutrinos.
(ii) The medium-energy region
($20 \rm{MeV} \lsim E_{\nu} \lsim 50 \, \rm{MeV}$)
exemplified by the higher energy region of supernova neutrinos.
(iii) The high-energy region ($50 \,\rm{MeV} \lsim E_{\nu}$)
represented by the solar flare neutrinos and atmospheric neutrinos,
whose energy can reach as high as $1 \sim 2$ GeV.
Neutrino-nucleus reactions
offer a great variety of ways to detect these astrophysical neutrinos.
Generally, the following types of
neutrino-nucleus reactions are possible:
\ben
\nu_{\ell} \, + \, ^ZA &\ra& \ell^- \, + \, ^{Z+1}A\, , \lab{eq:nuACC}\\
\bar{\nu_{\ell}}\, + \, ^ZA &\ra& \ell^+ \, + \, ^{Z-1}A\, ,
\lab{eq:nubarACC}\\
\nu_{\ell} \, + \, ^ZA &\ra& \nu_{\ell} \, + \, ^ZA\, ,  \lab{eq:nuANC}\\
\bar{\nu_{\ell}} \, + \, ^ZA &\ra& \bar{\nu_{\ell}} \, + \, ^ZA\, .
     \lab{eq:nubarANC}
\een
In the charged-current (CC) reactions,
eqs. (\ref{eq:nuACC}),(\ref{eq:nubarACC}),
the script $\ell$ is limited to the electron family
unless the incident energy is high enough to produce
a muon or a tau-lepton,
whereas the neutral-current (NC) reactions,
eqs.(\ref{eq:nuANC}),(\ref{eq:nubarANC}),
can always occur for $\ell=e,\,\mu$ and $\tau$.
The final nuclear state $A$ can in general be
either bound or unbound (if the latter is allowed energetically),
and all these will contribute to inclusive events
in which only the final leptons are monitored.
On the other hand, in a particular type of experiment
such as a radiochemical experiment,
only the contribution of particle-bound final states
will be registered.

To set the stage,
we first give a brief survey of the solar neutrino problem.
The Sun generates its energy primarily by changing protons
into $\alpha$-particles.
The actual reaction chain through
which $p+p+p+p \ra \,^4{\rm{He}}$ takes place
involves many intermediate nuclear reactions,
some of which are weak-interaction processes
emitting neutrinos.
The standard solar model (SSM) \cite{bah89,bu88}
gives a definite prediction
of the neutrino flux coming from
each of these neutrino-emitting reactions.
The first observation of the solar neutrinos
was achieved by Davis and his collaborators \cite{dav68}
through a radiochemical counting of $^{37}$Ar
produced in the reaction
$\nu_e + \,^{37}{\rm{Cl}} \ra e^- + \,^{37}{\rm{Ar}}$.
According to the SSM,
the $^{37}$Cl detector of Davis \etal's Homestake facility
should register $7.9 \pm 2.6$ SNU
(1 SNU $=\, 10^{-37}$ events per target atom per second),
but the observed value
is only $2.1 \pm 0.9$ SNU \cite{dav68}-\cite{dav90}.
This clear deficit of the neutrino flux
is the solar neutrino problem,
which is one of the most fundamental issues
in today's astrophysics.
An independent confirmation of the deficit
of the solar neutrino flux
has been obtained by the Kamiokande experiment
\cite{hiretal89,hiretal90,hiretal91},
in which the neutrino flux is monitored
through the leptonic reaction
$\nu_e + e^- \ra\,  \nu_e + e^-$
occurring in the large water Cerenkov counter.
The neutrino flux observed by the Kamiokande \cite{hiretal91} is
$0.46 \pm 0.05(\rm{stat}) \pm 0.06 (\rm{syst})$
times the SSM prediction,
corroborating the existence of the solar neutrino problem.
One thing to be noted here is
that the neutrino detection thresholds
of the Homestake and Kamiokande facilities
are rather high.
The Kamiokande with $E_{\rm{thresh}}=$ 7.5 MeV
is sensitive only to the neutrinos coming from
the reaction
$^{8}{\rm{B}} \ra \,^{8}{\rm{Be}} \, + \, e^+ \, + \, \nu_e$
($E_{\nu}^{\rm{max}}=15$ MeV),
whereas the chlorine detector is sensitive
to the $^{8}$B neutrinos
and, to a much lesser degree,
to the neutrinos coming from
$^{7}{\rm{Be}} + e^- \ra\, ^{7}{\rm{B}} + \nu_e$.
Since $^{8}$B is produced only towards the end
of the solar thermonuclear reaction chain,
the SSM prediction of the neutrino flux due to $^{8}$B
is extremely sensitive to the fine details of solar models.
By contrast,
the neutrino flux due to the primary
solar thermonuclear reaction
$p \, + \, p \, \ra\, ^2{\rm{H}} \, + \, e^+ \, + \, \nu_e$
reflects more directly the basic features of the SSM.
However, since the maximum neutrino energy for this process is
small ($E_{\nu}^{\rm{max}}$ = 0.420 MeV),
the detection of the pp-neutrinos requires
a neutrino absorption reaction with a very low Q-value.
The best target so far proposed for this purpose
is $^{71}$Ga.
The threshold of the
$\nu_e \, + \, ^{71}{\rm{Ga}} \, \ra \, e^- \, + \, ^{71}{\rm{Ge}}$
reaction is $E_{\rm{th}}=0.2332 \, \rm{MeV}$,
and the SMM prediction for the $^{71}$Ga target is
$132^{+20}_{-17}$ SNU \cite{bu88},
of which more than half is due to the pp neutrinos.
Two groups have recently published the results of Ga experiments.
The GALLEX collaboration \cite{GALLEX} reports
[$83 \pm 19\, (1\sigma)\, \pm 8$(syst) SNU],
and the SAGE collaboration \cite{SAGE1,SAGE2} gives
[$58^{+17}_{-24}\pm 14$(syst) SNU],
both significantly below the SSM value.
A wide range of possible solutions
to the solar neutrino problem have been proposed so far.
These ``solutions" fall into two categories.
In the first category (solar-model explanations)
a possible deviation from the SSM prediction
on the interior properties of the Sun is sought.
In the second category (particle-physics explanations),
the observed neutrino flux deficiency
is ascribed to unusual properties of the neutrinos.

Bahcall and Bethe \cite{bb93}
have recently carried out a systematic study
of solar-model explanations.
They studied a collection of 1000 precise solar models
in which each input parameter
(the principal nuclear reaction rates, the solar composition,
the solar age, and the radiative opacity)
for each model was drawn randomly from a normal
distribution with the mean value and standard deviation
appropriate to that variable.
None of these 1000 solar models
was compatible with the results
of the chlorine or the Kamiokande experiments.
Even if the solar models were artificially modified to reproduce
a $^8$B neutrino flux in agreement with the Kamiokande experiment,
none of these fudged models agreed with the chlorine observation.
Furthermore, the GALLEX and SAGE experiments
were found to disagree with the prediction of the solar models
by $2\sigma$ and $3\sigma$, respectively.
Thus, the totality of the existing solar neutrino data
seems to strongly indicate that the solar neutrino problem
has at least part of its origin in the hitherto unknown
properties of the neutrinos.

The particle-physics explanations invoke,
in one form or another, a mechanism
to convert electron neutrinos into
other particles (neutrino oscillation)
so that not all of the original $\nu_e$'s
produced in the core of the Sun
would reach the terrestrial detectors
(for a review, see e.g. \cite{kp89}).
The most promising particle-physics solution so far proposed
is the Mikheyev-Smirnov-Wolfenstein (MSW) effect
\cite{ms85,wol78},
which can be summarized as follows.
The three neutrino states,
$\nu_e$, $\nu_{\mu}$ and $\nu_{\tau}$,
are assumed to have different masses,
with the electron neutrino the lightest.
In the solar core with high electron densities,
the neutrinos acquire effective masses through
the neutrino-electron interactions.
The neutral-current interaction simply gives a universal shift
of the neutrino masses,
but the charged-current interaction,
which in the absence of muons and $\tau$-leptons
is operative only for $\nu_e$,
increases the $m_{\nu_e}$ selectively
so that, for sufficiently high electron densities,
the effective electron neutrino mass can become
heavier than those of the other neutrinos.
If the Hamiltonian involves a piece that causes
lepton-flavor mixing,
then in a critical region where
$m_{\nu_e}^{\rm{eff}} \sim m_{\nu_{\mu}}^{\rm{eff}}$,
or
$m_{\nu_e}^{\rm{eff}} \sim m_{\nu_{\tau}}^{\rm{eff}}$,
it is possible for $\nu_e$
to transform into the other neutrino species
via resonant transitions.
This phenomenon is called the MSW effect \cite{ms85,wol78}.
The parameters characterizing the MSW mechanism
are the neutrino mass differences in free space
and the strengths of flavor mixings.
By adjusting these parameters
one can vary the region of the electron neutrino spectrum
that is changed into the other neutrinos
through the resonant transitions.
The contemporary literature abounds in
articles that deal with the determination of the parameters
characterizing the MSW model or its variants.
At present, the bare neutrino mass difference of
about $10^{-2}$ to $10^{-3}$ eV
with appropriate corresponding mixing strengths
is compatible with the existing data.
However, this does {\it not} constitute unambiguous evidence
that the MSW mechanism is operative in the Sun;
the direct confirmation of the MSW effect
needs an additional set of experiments.
Neutrino-nucleus reactions are
of great importance in this context,
since in some favorable cases they allow us to identify
various types of neutrino reactions individually.
Of particular significance is the detection of neutral-current (NC)
reactions,
eq.(\ref{eq:nuANC});
since the NC reaction rate is the same
regardless of neutrino species,
a measurement of the NC reactions
gives the bolometric flux of all neutrino species
\cite{geretal62,dp79,che85}.
Therefore,
by combining information on the NC reactions
with that on the CC reactions,
one can obtain definitive information
on neutrino oscillation, or more specifically,
on the MSW effect.
In the planned Sudbury Neutrino Observatory
(SNO) \cite{ardetal87,ewa87},
which uses a 1,000-tonne, heavy-water Cerenkov counter,
one aims at the separate registration of
the CC and NC reactions on the deuteron
with the detection threshold as low as $E_{\nu}=$5 MeV.
The data obtained at the SNO will provide unambiguous
information on the neutrino flux for individual flavors
and consequently on the neutrino oscillation.
There are other nuclear targets
which are important in this context,
and we will discuss them later in the text.

The neutrino-nucleus reactions are not the only
place where neutral-current interactions appear.
The leptonic reactions:
\ben
\nu_{\ell} + e^-           &\ra & \nu_{\ell} + e^-\, ,  \lab{eq:nue}\\
\bar{\nu}_{\ell} + e^- &\ra& \bar{\nu}_{\ell} + e^- \, ,
   \lab{eq:nubare}
\een
where $\ell = e,\, \mu$ or $\tau$,
do involve the neutral-current contributions.
More specifically,
for $\ell = e$, the reactions can occur through
both the charged and neutral currents,
whereas for $\ell = \mu$ and $\tau$
only the neutral current will participate.
Therefore, water Cerenkov counters and liquid scintillation counters
in which these leptonic reactions provide main signals
do count both neutral-current (NC) and charge-current (CC) events.
It is to be noted, however, that
the NC and CC events
cannot be registered separately
by monitoring the leptonic reactions,
eqs.(\ref{eq:nue}),(\ref{eq:nubare}).
This situation makes the use of nuclear targets
particularly attractive.

We now turn to astrophysical neutrinos other than the solar
neutrinos.
The large water Cerenkov counters
at the Kamiokande and IMB
allowed the first observation of
the supernova neutrinos coming from SN1987A.
For this spectacular success
it was crucial that these facilities
had sufficiently high time resolutions
to clearly define supernova-related neutrino events.
Since all the existing nuclear-target detectors
use the radio-chemical counting method,
their time resolution is not sufficient
to allow the monitoring of supernova neutrinos.
In the near future, however,
counter experiments involving nuclear targets
will become possible.
The above-mentioned Sudbury Neutrino Observatory
(SNO) \cite{ardetal87,ewa87}
is a notable example.
As will be discussed in more detail later,
the SNO's ability to detect the NC reactions
with high time-resolution can be important
in the context of the supernova neutrinos as well.
We will also discuss some other nuclear targets
which can be very useful in connection with
supernova neutrino detection.

In considering the high-energy astrophysical neutrinos
[group (iii) in the above-described classification],
it is important to realize that,
as the incident neutrino energy increases,
the nuclear reactions become progressively more important
as compared to the leptonic processes.
This implies that even the water Cerenkov counter and
liquid scintillation counter,
which for low-energy neutrinos
primarily detect the electrons from the leptonic reactions,
will start detecting the nuclear reactions
\cite{hax87}.
As will be discussed in the following,
this offers yet another interesting use of these facilities.
A notable example is the water Cerenkov counter measurement
of the ratio of the muonic neutrino flux to
the electronic neutrino flux in the atmospheric neutrinos.

One of the fundamental prerequisites for nuclear targets
envisaged as neutrino detectors is that
reliable estimates of the relevant reaction cross sections
be available.
This presents an interesting and challenging task
to nuclear physicists.
\footnote{The cross sections of the leptonic processes,
eqs.(\ref{eq:nue}),(\ref{eq:nubare}),
by contrast are accurately known
from the standard electroweak theory.
There is in fact one particular type of
neutrino-nucleus reaction for which the cross section
can be determined mode-independently;
for the coherent scattering
$\nu + A({\rm{gnd}}) \ra \nu + A({\rm{gnd}})$,
the cross section is determined
by the number of nucleons in $A$.
The coherent process plays an important role
in dark matter search \cite{sadetal88}
as well as in certain types of
solar neutrino experiments \cite{ds84,ckw85,mar87},
but we will not discuss the coherent scattering here.
For a review of this topic, see \cite{bah89}.}
The great variety of relevant nuclear targets
and the wide energy range of the astrophysical neutrinos
to be considered necessitate the use of many different approaches
that can be found in the nuclear physics arsenal.
In this review,
we survey the current status of theoretical attempts
at obtaining reliable estimates of the neutrino-nucleus reactions
of astrophysical interest.
To highlight the basic aspects of
various approaches so far used and their possible limitations,
we shall organize our topics
according to the calculational methods involved.
We describe in chapter 2 the direct microscopic calculation
which is relevant to the lightest nuclei.
In chapter 3 we discuss a model-independent determination
of the cross sections
based on the elementary-particle treatment.
Chapter 4 deals with the empirical
effective operator method.
The use of (p,n) reactions
as calibrators of the Gamow-Teller matrix elements
is surveyed in chapter 5.
Chapters 6 and 7 are concerned with high-energy
neutrino-nucleus reactions.
The Fermi gas model and its possible refinement
are discussed in chapter 6,
while the treatment of semi-inclusive
reaction is addressed in chapter 7.
Additional remarks are made in chapter 8.

\newpage

\noin
{\bf Chapter 2 \hspace{1cm}
``Ab initio" calculations based on realistic wave functions}

In the so-called realistic description of nuclei one works with
the nuclear Hamiltonian
\beq
H_N = \sum_{i=1}^{A} T_i + \sum_{i,j}^{A} V_{ij}, \lab{eq:Hnucl}
\eeq
where $A$ is the mass number,
$T_i$ is the nucleon kinetic energy,
and $V_{ij}$ is the realistic nucleon-nucleon potential.
Although $H_N$ is called the realistic Hamiltonian
due to the fact that $V_{ij}$
reproduces the observed properties of
the two-nucleon systems reasonably well,
$H_N$ is an effective Hamiltonian
of highly operational nature.
To arrive at $H_N$
starting from the fundamental QCD picture
is far from trivial.
First, the dynamical degrees of freedom of quarks and gluons
need to be translated into
the effective degrees of freedom of hadrons.
Second, the Hilbert space of hadrons must be truncated
down to those of non-relativistic nucleons
interacting via potentials.
Despite these fundamental problems,
it is generally believed
that solutions for the Schr\"{o}dinger equation
with the Hamiltonian $H_N$
give reasonably reliable nuclear wave functions,
and it is usually considered
as an {\it ab initio} approach
to calculate matrix elements of
nuclear electromagnetic or weak-interaction processes
with the use of these wave functions
along with the standard non-relativistic reduction
of the single-nucleon responses to
the relevant currents \cite{mrr69,bli73}.
Two major problems encountered
in this {\it ab initio} approach are as follows.
First, it is in general extremely difficult to obtain
an exact solution of the $A$-body Schr\"{o}dinger equation
$H_N\Psi(1,2, \ldots A) = E\Psi(1,2, \ldots A)$,
and hence one is usually forced to work with rather drastically
truncated model wave functions $\Psi_0$;
shell-model wave functions are the most commonly used
and most successful example.
Now, if the matrix element of a nucleon observable ${\cal O}_N$
is calculated using model wave functions
$\Psi_0^i$ and $\Psi_0^f$, one expects
$<\Psi^f|{\cal O}_N|\Psi^i> \neq <\Psi_0^f|{\cal O}_N|\Psi_0^i>$.
This difference is called core-polarization effect.
Another point is
that eliminating all hadronic degrees of freedom
but those of the nucleons is an approximation
and, even if one can calculate
$<\Psi^f|{\cal O}_N|\Psi^i>$ exactly, it may still differ
from the corresponding true matrix element.
One usually accounts for this difference
by adding to ${\cal O}_N$ many-body operators
${\cal O}^{\prime}$,
which represent the effects of the
eliminated hadronic degrees of freedom.
These additional terms are called
the exchange-current operators.

Now, to calculate the core-polarization
and exchange-current effects
in complex nuclei is one of the most challenging
problems in nuclear physics and,
although many important achievements have so far been made
in this domain,
the problem is far from solved.
However, nuclei with small mass numbers ($A \leq 4$)
offer important exceptions.
For these lightest nuclei
one can obtain sufficiently ``realistic" wave functions
either by directly solving the Schr\"{o}dinger equation
or through variational calculations,
eliminating thereby the core-polarization problem.
For these systems, therefore,
one can carry out {\it ab initio} calculations
of the transition matrix elements
of electromagnetic and weak-interaction processes,
provided a reasonably reliable method to incorporate
the exchange-current effects exists.

\vspace{0.4cm}
\noin
2.1. \hspace{0.3cm} $\nu-d$ reactions

The A=2 systems are the simplest nuclei,
for which one can obtain
reasonably reliable wave functions
corresponding to the realistic Hamiltonian eq.(\ref{eq:Hnucl}).
This allows us to estimate with high accuracy
the cross sections of the neutrino-deuteron reactions.
These cross sections are important
for deriving useful astrophysical information
from the experimental data that will soon become available
at the Sudbury Neutrino Observatory (SNO) \cite{ardetal87,ewa87}.
The neutrino-deuteron reactions that are relevant to the SNO are:
\ben
\nu + d            &\ra& \nu' + n + p, \lab{eq:nudNC}\\
\bar \nu + d    &\ra& \bar \nu + n + p. \lab{eq:nubardNC}\\
\nu_e + d         &\ra& e^- + p + p, \lab{eq:nudCC}\\
\bar \nu_e + d &\ra& e^+ + n + n, \lab{eq:nubardCC}
\een
As mentioned earlier,
one of the great advantages of the SNO is its capability
to detect both the neutral- and charged-current reactions
and record them separately,
and this is crucially important \cite{che85}
for proving that the MSW effect \cite{ms85} is
indeed a solution for the solar neutrino problem \cite{bah89}.

The earlier calculations
\cite{eb68,nkkk,bkn87}
were mainly meant for low-energy solar neutrinos
and therefore included only s-wave
for the final N-N relative motion.
Extended calculations that take account of higher partial
waves and, correspondingly, the contributions of
the forbidden-type transitions
were carried out by Ying, Haxton and Henley \cite{yhh}
(to be referred to as YHH1)
and by Tatara, Kohyama and Kubodera \cite{tkk}
(to be referred to as TKK)
up to $E_{\nu}=$ 170 MeV.
Unfortunately, the cross sections obtained
in these two detailed calculations
show alarmingly large discrepancies for higher energies;
the YHH1 results become significantly larger than those of TKK
around $E_{\nu} \sim 20$ MeV
and the cross section ratios reach
$\sim2$ towards $E_{\nu}= 170$ MeV.
To clarify the origin(s) of these discrepancies,
Doi and Kubodera \cite{dk92a} (to be referred to as DK1)
carried out a systematic comparison
of the formalisms used in YHH1 and TKK.
The numerical results of the independent calculation
in DK1 \cite{dk92a}
suggested possible numerical errors in YHH1,
and this suggestion was confirmed
by the authors of YHH1.
The results of a revised estimation by Ying, Haxton and Henley
\cite{yhh92} (to be referred to as YHH2)
are in essential agreement with those of TKK and DK1.

Beyond the $\sim$20 \% level precision, however,
there still exist discrepancies among
the results of TKK, DK1 and YHH2.
The cross sections in TKK
begin to be systematically smaller
than the corresponding numbers
in DK1 and YHH2 as $E_{\nu}$ surpasses $\sim 100$ MeV,
and the differences for some channels reach
$\sim$20 \%
towards $E_{\nu}=$ 170 MeV.
It has been demonstrated in DK1
that this discrepancy arises
because TKK dropped the term quadratic in
$\mbox{\bm $J$}^V$,
the space component of the vector current.
At low energies,
the contribution of $\mbox{\bm $J$}^V$
is known to be much less important
than that of the time-component $\rho_V$.
TKK, assuming that this behavior would persist
for higher energies,
ignored the quadratic term $|\mbox{\bm $J$}^V|^2$
that appears in squaring the transition matrix element
to obtain the cross section.
According to DK1, however,
the contribution of $\mbox{\bm $J$}^V$ is comparable to
or even larger than that of $\rho_V$
for $E_{\nu} \gsim 10$ MeV,
and ignoring $|\mbox{\bm $J$}^V|^2$ leads to
an underestimation of the cross sections by up to $\sim$20 \%.
Meanwhile,
$\sigma(\bar \nu_e + d \ra e^+ + n + n)$ of DK1
becomes progressively larger than the corresponding results
in TKK as $E_{\nu}$ decreases from 7 MeV;
e.g., at $E_{\nu}=6$ MeV,
$\sigma(\bar \nu_e + d \ra e^+ + n + n)_{\rm{DK1}}/
\sigma(\bar \nu_e + d \ra e^+ + n + n)_{\rm{TKK}}=1.07$.
(At this low energy the contribution of $|\mbox{\bm $J$}^V|^2$
ignored in TKK is completely negligible.)
A recent study by Doi and Kubodera
\cite{dk92b} (to be referred to as DK2)
indicates that this behavior is caused by
the approximate Q-value expression used for this channel in DK1.
As far as comparison with YHH2 is concerned,
we remark that
the exchange-currents effects are altogether ignored in YHH2.
This is estimated \cite{tkk} to lead
to a general underestimation of
the cross sections by up to $\sim$5 \%.
Thus each of the published works,
TKK \cite{tkk}, DK1 \cite{dk92a} and YHH2 \cite{yhh92},
has some point(s) to be improved upon.

Kohyama and Kubodera \cite{kk92}
have extended the calculation of Tatara \etal. \cite{tkk}
by including the contributions of the $|J_V|^2$ terms.
The results of Kohyama and Kubodera \cite{kk92}
and those of DK2 \cite{dk92b}
have been found to agree with each other within 2 \%.
Since these two calculations are independent,
both in the calculational methods
and in choosing various input parameters,
it is reassuring that their results agree to this degree.
(According to TKK, different choices of the nuclear potential
from among realistic nucleon-nucleon interactions
affects the cross sections by up to 1 \%.)
We tabulate in table 1 in Appendix
the cross sections for the reactions eqs.(8)-(11)
obtained by Kohyama and Kubodera \cite{kk92};
5 \% uncertainties are attached to the cross sections.
Table 1 represents the best available estimates of
the neutrino-deuterium cross sections
up to $E_{\nu}=170$ MeV.

One generally expects that the ratios
$R_{\nu} \equiv
\sigma(\nu_e + d \ra e^- + p + p)
   /\sigma(\nu + d \ra \nu' + n + p)$
and
$R_{\bar{\nu}} \equiv
\sigma(\bar \nu_e + d \ra e^+ + n + n)
   /\sigma(\bar \nu + d \ra\bar \nu + n + p)$
are much less sensitive to the nuclear models
than the absolute cross sections themselves.
Indeed, the $R_{\nu}$ and $R_{\bar{\nu}}$
obtained by Kohyama and Kubodera \cite{kk92}
agree with the corresponding values given in DK2 \cite{dk92b}
within 0.2 \%.
Therefore,
it seems reasonably safe
to assign 1 \% uncertainties to the calculated values
of $R_{\nu}$ and $R_{\bar{\nu}}$.

The average cross section
for the electronic neutrinos from the $\mu^+$ decay
calculated using the results of \cite{kk92} is:
$<\sigma(\nu_e + d \ra e^- + p + p)>_{\mu^+\rm{-decay}}
=5.4 \times 10^{-41}\, \rm{cm}^2$.
This is consistent with the observed value
$\sigma=(5.2 \pm 1.8) \times 10^{-41}\, \rm{cm}^2$
\cite{wiletal80}.

Apart from testing the MSW effect for the solar neutrinos,
Bahcall \etal. \cite{bkn87}
considered the merit of the SNO in determining the masses of
heavy-flavor neutrinos through arrival-time-delay measurements
of supernova neutrinos.
Some other possible uses of the SNO
for solar and supernova neutrinos
were discussed by Balantekin and Loreti \cite{bl92}.
The SNO can also be used to study neutrinos
of extremely high energies such as solar-flare
and atmospheric neutrinos.
The above described ``{\it ab initio}" calculation
probably can be extended up to
$E_{\nu} \sim $ several hundred MeV,
but its reliability will diminish rather significantly
for these high energies.
At yet higher energies,
the results are expected to approach
those of the high-energy approximation to be discussed
in chapter 6.

\vspace{0.4cm}
\noin
2.2. \hspace{0.3cm} Neutrino-production reaction
on the lightest nuclei

Although our review is primarily concerned
with neutrino-nucleus reactions,
this is a good place to discuss the latest development
in the theoretical treatment of
neutrino-production reactions on the lightest nuclei.
As mentioned earlier, the pp reaction
\beq
p+p \ra d + e^+ + \nu_e\,\, , \lab{eq;pp}
\eeq
is the primary solar thermonuclear reaction
and produces a dominant part of the solar neutrino flux \cite{bah89}.
The ``{\it hep}" reaction
\beq
^{3}\rm{He} + p \ra\, ^{4}\rm{He} + e^+ +\nu_e\,\, , \lab{eq;Hep}
\eeq
contributes only a tiny portion
of the solar neutrino flux \cite{bah89},
but the maximum neutrino energy of the {\it hep} process
($E_{\nu}^{\rm{max}}=18.77$ MeV)
is the largest of all the solar neutrinos.
Carlson \etal. \cite{crsw91}
have carried out a detailed {\it ab initio} calculation
of the cross sections for these reactions,
using as a nucleon-nucleon interaction
the Argonne $v_{14}$ potential.
To reduce the uncertainty in the axial exchange current operator,
its matrix element has been adjusted so as to reproduce the
measured Gamow-Teller matrix element for tritium $\beta$-decay.
For the pp reaction,
the calculated astrophysical $S$ factor
and its derivative at zero energy are:
$S(E=0)=4.00 \times 10^{-25}$ MeV b, and
$dS(E)/dE|_{E=0}=4.07(1 \pm 0.051) \times 10^{-25}$ MeV b;
these are close to the values quoted in \cite{bu88}.
Carlson \etal. find that exchange-current contribution
enhances the capture rate by 1.5 \%.
This is considerably smaller than
the enhancement found in earlier works \cite{gh72,drr76}.
In the absence of Coulomb distortion
the transition matrix element of the pp reaction
is just the hermitian conjugate of that of the
$\nu_e + d \ra e^- + p + p$ reaction.
For the thermal pp reaction, however,
the Coulomb repulsion between the two protons
is extremely important and responsible for
the reduction of exchange-current contribution
as compared with the typical values found
for $\nu_e + d \ra e^- + p + p$ reaction \cite{tkk}.

As for the $^3$He$(p,e^+ \nu)^4$He reaction rate,
Carlson \etal. \cite{crsw91} obtained
$S(E=0)=1.3 \times 10^{-23}$ MeV b,
which is smaller by a factor of $\sim 6$ than
the values obtained earlier by Werntz and Brennan \cite{wb73}
and by Tegner and Bargholtz \cite{tb83}.
This change is caused primarily by the exchange current effect,
which cancels the impulse approximation contribution.
This is one of the most dramatic examples
of the exchange-current effect.

\newpage

\noin
{\bf Chapter 3 \hspace{1cm}
Elementary-particle treatment (EPT)}

The elementary-particle treatment (EPT)
was first introduced by Kim and Primakoff \cite{kp65}
and by Fujii and Yamaguchi \cite{fy64}.
In EPT, instead of describing nuclei in terms of
nucleons or other constituents,
one treats nuclei as ``elementary" particles
with given quantum numbers.
A transition matrix element
for a given process
is parametrized in terms of the {\it nuclear} form factors
solely based on the transformation properties
of the relevant current and nuclear states
\cite{kp65,gp69,del72}.
Insofar as all of these nuclear form factors
can be deduced empirically from experimental data,
one can make a totally model-independent prediction
on every observable for the given transition.
In fact, this requirement is an extremely stringent one,
and so far the EPT calculation has been carried out
completely
only for the triad of $^{12}$B -$^{12}$C -$^{12}$N.

This particular $A$=12 case, however, deserves special attention.
One reason is that
the $^{12}$C nucleus, which is abundantly contained
in ordinary liquid scintillators,
can be a very important nuclear target in neutrino astrophysics.
Although the threshold energies of the neutrino reactions
on $^{12}$C are too high for solar neutrino detection,
they are low enough for observing energetic neutrinos
originating from stellar collapses.
The neutrino energies up to $E_{\nu} \sim 100$ MeV
are expected to be of relevance for stellar collapse neutrinos.
Another reason is
provided by the intensive programs
of beam-dump neutrino experiments
at Los Alamos \cite{alletal90,kraetal92,koeetal92,freetal93}
and the Rutherford Laboratory \cite{dreetal90,zei85}.
Some of these experiments directly aim at observing
the neutrino -$^{12}$C reactions.
In other particle-physics experiments \cite{alletal90}
main interest is in the $\nu-e$ scattering
but here also,
to eliminate the dominant background
due to the neutrino -$^{12}$C reaction,
one needs to know its cross sections with reasonable accuracy.

The processes of relevance here
are the superallowed neutral-current (NC) reactions
\ben
\nu + \,^{12}{\rm{C}}(0^+;{\rm{gnd}}) & \ra &
\nu' + \,^{12}{\rm{C}}^*(1^+;15.1 {\rm{MeV}}), \lab{eq:NuC12NC}\\
\bar \nu + \,^{12}{\rm{C}}(0^+;{\rm{gnd}}) & \ra &
\bar \nu' + \,^{12}{\rm{C}}^*(1^+;15.1\rm{MeV}), \lab{eq:NubarC12NC}
\een
and the superallowed charged-current (CC) reactions
\ben
\nu + \,^{12}{\rm{C}}(0^+;{\rm{gnd}}) & \ra &
e^- + \,^{12}{\rm{N}}(1^+;{\rm{gnd}}), \lab{eq:NuC12CC}\\
\bar \nu + \,^{12}{\rm{C}}(0^+;{\rm{gnd}}) &\ra&
e^+ + \,^{12}{\rm{B}}(1^+;{\rm{gnd}}). \lab{eq:NubarC12CC}
\een
The final nuclear states of the above reactions
form a triad of $J^P=1^+, T=1$ states.
The existence of these superallowed transitions
is favorable for simultaneous monitoring
of the NC- and CC-reactions.
(The elastic scattering leading to
$^{12}{\rm{C}}(0^+;\rm{gnd})$
is unobservable in normal circumstances.)
Fukugita, Kohyama and Kubodera \cite{fkk88}
showed that EPT allows
the completely model-independent determination
of the cross sections for the reactions
eqs.(\ref{eq:NuC12NC})-(\ref{eq:NubarC12CC})
for $E_{\nu} \lsim$ 100 MeV.
A similar treatment was presented by Mintz and Pourkaviani
\cite{mp89,pm90}.
Although the use of EPT is at present limited
to the transitions listed above,
this is not an obstacle for monitoring these exclusive processes,
as discussed in \cite{fkk88}.
Here we summarize the calculation of
Fukugita \etal. \cite{fkk88}.

The effective Hamiltonian responsible
for the NC reactions is given by
\beq
H_{eff}=G_F/ \sqrt{2} \, J^3_{\mu}L^3_{\mu}, \lab{eq:Hnc}
\eeq
where $G_F$ is the Fermi constant,
$L^3_{\mu}=\bar \psi_{\nu} \gamma_{\mu} (1-\gamma_5)
\psi_{\nu}$,
and $J^3_{\mu}=(1-2\sin^2\theta_W)V^3_{\mu} + A^3_{\mu}$,
with $\theta_W$ being the Weinberg angle.
(The isoscalar neutral current, which cannot
cause nuclear excitations, has been dropped.)
The effective Hamiltonian
for the CC reactions is given by
\beq
H_{eff}=[G_F \, \cos\theta_c/ \sqrt{2}]\,
J_{\mu}^{(\pm)}L_{\mu}^{(\mp)},
\eeq
where $\theta_c$ is the Cabibbo angle;
$L_{\mu}^{(-)}=\bar \psi_e \gamma_{\mu}(1-\gamma_5)
\psi_{\nu}$,
and
$L_{\mu}^{(+)}=L_{\mu}^{(-) \dagger}$;
$J_{\mu}^{(\pm)}=V_{\mu}^{(\pm)} + A_{\mu}^{(\pm)}$.
The most general form of the matrix elements
of the vector and axial-vector currents
for the NC reactions is
\beq
<^{12}{\rm{C}}(1^+);p_b|
V_{\mu}^3(0)|^{12}{\rm{C}}(0^+);p_b>
 = \frac {1}{2}
\frac {\epsilon_{\mu \nu \lambda \sigma}q_{\nu}
Q_{\lambda}\xi_{\sigma}F_M(q^2)}{2M}
\eeq
\beq
<^{12}{\rm{C}}(1^+);p_b|
A_{\mu}^3(0)|^{12}{\rm{C}}(0^+);p_b>
 = \frac {1}{2} [\xi_{\mu}F_A(q^2) + q_{\mu}(\xi \cdot q)
                  \frac {F_P(q^2)}{m_{\pi}^2}
                 +Q_{\mu}(\xi \cdot q) \frac {F_T(q^2)}{2M}],
\eeq
where $q=p_b -p_a$, $Q=p_b+p_a$, $\xi$ is
the polarization vector of the spin-1 nucleus,
and $M=(M_a+M_b)/2$ is the average of
the initial and final nuclear masses.
All information on nuclear dynamics is contained
in the {\it nuclear} form factors,
$F_M(q^2)$, $F_A(q^2)$, $F_P(q^2)$, and $F_T(q^2)$.
Here the nuclear form factors are classified
in the ``cartesian" representation.
An alternative formalism is described in \cite{gp69,bp79}.
For the CC reactions the Wigner-Eckart theorem
in isospin space leads to
\beq
<^{12}{\rm{N}}(1^+);p_b|
V_{\mu}^{(+)}(0)|^{12}{\rm{C}}(0^+);p_b>
 =  \frac {\epsilon_{\mu \nu \lambda \sigma}q_{\nu}
   Q_{\lambda}\xi_{\sigma}F_M(q^2)}{2M}
\eeq
\beq
<^{12}{\rm{N}}(1^+);p_b|
A_{\mu}^{(+)}(0)|^{12}{\rm{C}}(0^+);p_b>
 =  \xi_{\mu}F_A(q^2) + q_{\mu}(\xi \cdot q)
                  \frac {F_P(q^2)}{m_{\pi}^2}
                 +Q_{\mu}(\xi \cdot q) \frac {F_T(q^2)}{2M}.
\eeq
Now, the contribution of the $F_P$ term
to the observables is proportional to
$(\rm{lepton \,\, mass})^2$
and therefore can be dropped in the present energy regime
($E_{\nu} \lsim 100$ MeV)
for which muon-production can be ignored.
For the remaining nuclear form factors
one can first consider their values at $q^2=0$,
and examine their $q^2$-dependence later.
Define
$F_M \equiv F_M(q^2=0)$, $F_A \equiv F_A(q^2=0)$,
and $F_T \equiv F_T(q^2=0)$.
$F_M$ can be determined
from the width of the $\gamma$-decay
$^{12}{\rm{C}}^*(1^+;15.1{\rm{MeV}})
\ra \,^{12}{\rm{C}}(0^+;{\rm{gnd}})$,
and the CVC.
{}From $\Gamma_{\gamma}^{exp}=38.5 \pm 0.8$ MeV
\cite{ajz85},
we obtain
$F_M=(1.516 \pm 0.016) \times 10^{-3} \, \rm{MeV}^{-1}$.
$F_A$ and $F_T$ can be determined from the $\beta^+$ decay
$^{12}{\rm{N}}(1^+;{\rm{gnd}})
\ra \,^{12}{\rm{C}}(0^+;{\rm{gnd}})$
and the $\beta^-$ decay
$^{12}{\rm{B}}(1^+;{\rm{gnd}})
\ra \,^{12}{\rm{C}}(0^+;{\rm{gnd}})$.
The $ft$-value of the $\beta$-decay is written as
\beq
ft = \frac {2\pi^2 \ln 2}
{m_e^5(G_F \cos\theta_c)^2F_A^2(1+\frac{2}{3}E_0F_T/F_A)},
\lab{eq:ft12}
\eeq
with $E_0$ the maximum $\beta$-particle energy.
The measurement of the $\beta$-ray angular distribution from
aligned $^{12}$B and $^{12}$N parent nuclei
gives the angular correlation quantities $\alpha_{\pm}$,
which are related to $F_T$ as
\cite{moretal76,hp77,gre85}.
\beq
\alpha_+ + \alpha_- = - \frac{4}{3}F_T/F_A.
\eeq
The experimental values \cite{masetal79,lebetal78},
$\alpha_{-}=(0.02 \pm 0.12) \times 10^{-3}\,\rm{MeV}^{-1}$ and
$\alpha_{+}=-(2.72 \pm 0.27) \times 10^{-3}\,\rm{MeV}^{-1}$,
give $F_T/F_A=(2.01 \pm 0.23) \times 10^{-3}\,\rm{MeV}^{-1}$.
Then the contribution of the $F_T$ term in eq.(\ref{eq:ft12}) is
$(2/3)E_0(F_T/F_A)=0.02$ for $E_0=15$ MeV.
The importance of this term relative to 1
being smaller than typical isospin-symmetry breaking effects
within nuclei ($\alpha Z \sim 6\%$),
the $F_T$ term in eq.(\ref{eq:ft12}) can be ignored.
Experimentally \cite{ajz85},
$\log (ft)=4.067 \pm 0.002$ for
$^{12}{\rm{B}} \ra \,^{12}{\rm{C}}$, and
$\log (ft)=4.120 \pm 0.003$
for $^{12}{\rm{N}} \ra \,^{12}{\rm{C}}$.
The difference between $(ft)^+$ and $(ft)^-$ gives a measure
of the isospin symmetry breaking quoted above
(insofar as there is no second-class currents).
Using $ft=12444 \pm 830$ s,
which is the average of $(ft)^+$ and $(ft)^-$
with the error corresponding to their difference,
we obtain $F_A=0.711 \pm 0.024$.

As $E_{\nu}$ grows,
one must consider the $q^2$-dependence
of the nuclear form factors.
The total cross section for the NC reactions
including the $q^2$-dependence is given by
\ben
\sigma   &=& \sigma_0F_A^2E_{\nu'}I, \\
\sigma_0 &=& 3G_F^2/2\pi = 2.53 \times 10^{-44}{\rm{cm}}^2
/(\rm{MeV})^2,\\
I &=&  \frac{1}{2} \int_{-1}^{1}dz f(\mbox{\bm $q$}^2)
   [(1- \frac {z}{3}) \pm \frac{4}{3}(E_{\nu}+E_{\nu'})
       (1-2\sin^2\theta_WW) (1-z) \frac {F_M}{F_A} \non \\
         & &   +\,\frac {2}{3} [E_{\nu'}E_{\nu}(1-z^2) +
      (1-z) \mbox{\bm $q$}^2] (1-2\sin^2\theta_W)^2
             (\frac {F_M}{F_A})^2 \non \\
    & &  -\frac {2}{3} \Delta M (1+z) \frac {F_T}{F_A}
   +\frac {1}{3} \mbox{\bm $q$}^2(\frac {F_T}{F_A}^2) \,],
\een
where the upper (lower) sign refers
to the neutrino (anti-neutrino) reaction;
$z=\cos \theta$ with $\theta$ being the neutrino scattering angle;
$E_{\nu'}=E_{\nu}-\Delta M$ with $\Delta M= 15.1$ MeV;
$f(\mbox{\bm $q$}^2) = [F_A(q^2)/F_A]^2$
with
$q^2 = (\Delta M)^2 - \mbox{\bm $q$}^2$.
The cross sections for the CC reactions are given  by
\beq
\sigma=2\sigma_0 \cos \theta_c p_e E_eF^{\pm}(Z_f;E_e)F_A^2I,
\eeq
where the upper (lower) sign refers to the neutrino
(anti-neutrino) reaction.
Here $I$ is as defined previously except the replacements
$E_{\nu'} \ra E_e$, and
$(1-\sin^2\theta_W) \ra 1$;
$F^{\pm}(Z_f;E_e)$ denotes the Fermi function
for the Coulomb correction.
$E_e = E_{\nu}-\Delta M$,
with $\Delta M = 16.827$ MeV ($13.880$ MeV) for
$^{12}{\rm{C}} \ra \,^{12}{\rm{N}}$
($^{12}{\rm{C}} \ra \,^{12}{\rm{B}}$).
The above expression has been obtained
by assuming that the $q^2$-dependences of
$F_M(q^2)$ and $F_T(q^2)$ are the same as that of $F_A(q^2)$.
{}From an EPT analysis of
the $^{12}{\rm{C}}+ \mu^-
\ra \,^{12}{\rm{B}} + \nu_{\mu}$ process,
Nozawa \etal. \cite{nkk83} showed that
$F_A(q^2)/F_A = F_M(q^2)/F_M$
holds within a 10 \% accuracy at least up to
$q^2 = 0.74 m_{\mu}^2$.
For $F_T(q^2)$ no direct experimental information is available
but, since its contribution is minor, its precise form
has little significance.
Now,
$f(\mbox{\bm $q$}^2) \cong [F_M(q^2)/F_M]^2$
can be determined model-independently
from $F_M(q^2)$ obtained from inelastic electron scattering:
$e +\,^{12}{\rm{C}}
\ra e^{\prime} + \,^{12}{\rm{C}}^*(1^+;15.1 \, {\rm{MeV}})$.
According to \cite{cheetal73},
\beq
f(\mbox{\bm $q$}^2)=
[1-(1-\rho)/6(b|\mbox{\bm $q$}|)^2]^2
\exp[-\frac {1}{2}(b|\mbox{\bm $q$}|)^2],
\eeq
where $b=1.881 \pm 0.053$ fm, and
$\rho=0.23 \, (1 \pm 0.1)$.

Using the EPT summarized above,
Fukugita \etal. \cite{fkk88}
were able to determine the cross sections
for eqs. (\ref{eq:NuC12NC})-(\ref{eq:NubarC12CC})
up to $E_{\nu} \sim 100$ MeV
within $\sim 12 \%$ accuracies;
these uncertainties reflect the ambiguities in the input data
and isospin symmetry breaking in nuclei.
Similar results were obtained
by Mintz and Pourkaviani \cite{mp89,pm90}.
As in the A=2 case discussed earlier,
the ratio of the NC cross section to the CC cross section
is expected to be less affected
by the nuclear physics ambiguities.
When we vary the nuclear form factors
within $\sim$10 \% (which are typical errors given in ref. \cite{fkk88}),
$R \equiv
\sigma[^{12}{\rm{C}}(\nu_e,\nu_e^{\prime})
^{12}{\rm{C}}^*(1^+;15.1{\rm{MeV}})] /
\sigma[^{12}{\rm{C}}(\nu_e,e^{-})^{12}{\rm{N}}({\rm{gnd}})]$
is found to change by less than $\sim$2 \%.

We now remark
on another semi-empirical method developed
by Walecka and Donnelly
to estimate the matrix elements of
weak-interaction transitions in light nuclei.
In the Walecka-Donnelly (W-D) method
\cite{wal75,dw75,dp79},
one assumes that the weak-interaction transition operator
can be approximated
by a sum of single-particle multipole operators
that arise in the impulse approximation
and that the electomagnetic data can be used
to place constraints on the nuclear matrix elements
of these operators.
Because of these basic assumptions
it should be obvious that,
contrary to the often-made claim,
the W-D method is {\it not} a model-independent approach.
Apart from the fundamental assumption
pertaining to the impulse approximation,
an important question is to what extent
one can identify the effective matrix element
of a single-nucleon operator ${\cal O}_N$
appearing in the weak-interaction process
with that of the ${\cal O}_N$ that
appears in the electromagnetic process.
Chemtob and Rho \cite{cr71}
emphasized that the exchange-currents for the
electromagnetic interaction can be drastically
different from those of the weak processes.
This point has been reemphasized in \cite{crsw91}.
The EPT is free from these problems and hence,
from a formal point of view,
definitely a preferable approach.
Having said this,
we must quickly add that EPT
in its rigorous form can be used at present
only for the very limited case of A=12.
By contrast, the W-D method,
owing to the additional assumptions made on the
possible form of the transition operators,
can be applied to many cases
where the lack of necessary experimental information
renders EPT at moment powerless.
Thus, in practice, the W-D method
is a very useful approach insofar as
some independent information is available
to check its reliability for given individual cases.
The use of experimental data
to constrain the parameters involved
in the W-D method is in general expected to
reduce nuclear-model dependence significantly,
but the remaining possible uncertainties should
be kept in mind.

Donnelly and Peccei \cite{dp79} used the W-D method
to estimate the cross sections for
the NC reactions eqs. (\ref{eq:NuC12NC}),(\ref{eq:NubarC12NC}).
While a good agreement is seen
between Donnelly and Peccei's
results and those of Fukugita \etal. \cite{fkk88}
at lower incident energies,
disagreement becomes apparent for $E_{\nu} \gsim 25$ MeV,
and it amounts to 60 \% at $\sim 50$ MeV.
Furthermore, while Fukugita \etal. report
an appreciable difference between
$^{12}{\rm{C}}(\nu, \nu')^{12}{\rm{C}}^*$ and
$^{12}{\rm{C}}(\bar \nu, \bar \nu')^{12}{\rm{C}}^*$
at higher energies
due to the interference term $F_AF_M$,
this behavior is not reproduced in \cite{dp79}.
These discrepancies seem to signal the breakdown
of some assumptions involved in the W-D method
\cite{dp79} as applied to higher incident neutrino energies.
On the other hand, for low neutrino energies,
the EPT results justify the W-D method \cite{wal75},
in which the leading-order impulse-approximation
transition  matrix element $<\!\mbox{\bm $\tau \sigma$}\!>$
in the weak interaction process is deduced
from the observed M1 strength.
For $^{12}{\rm{C}}(\nu_e, e^-)^{12}{\rm{N}}$
the estimate of Fukugita \etal. \cite{fkk88}
is significantly smaller
than the published result of the W-D method \cite{don73};
specifically, for the cross section
averaged over the $\nu_e$ spectrum
due to the $\mu$-decay,
the EPT gives
$<\sigma[^{12}{\rm{C}}(\nu_e, e^-)^{12}{\rm{N}}({\rm{gnd}})]>
=9.2 \times 10^{-42}\, {\rm{cm}}^2$,
which is about 25 \% lower than that of Donnelly \cite{don73}.
However, a more recent result of the W-D method \cite{don92}
agrees with the EPT result.

The \underline{KA}rlsruhe-\underline{R}utherford
\underline{M}edium \underline{E}nergy
\underline{N}eutrino (KARMEN) experiment
has been intensively pursued at the
spallation facility of the Rutherford Laboratory
\cite{dreetal87,zei85,dreetal90}.
The pulsed beam dump neutrino source
provides monoenergetic $\nu_{\mu}$
with $E_{\nu}=29.8$ MeV from $\pi^{+}$ decay at rest
as well as $\nu_e$ and $\bar{\nu}_{\mu}$
with energies up to 52.8 MeV
coming from the subsequent $\mu^{+}$ decay.
A 56 t liquid scintillation calorimeter \cite{dreetal90}
is used as massive live target of $^{12}$C.
The NC reaction eq.(\ref{eq:NuC12CC}) has been observed
\cite{bodetal91}
by monitoring the $\gamma$ emission
back to the ground state
($E_{\gamma}=15.1$ MeV, B.R = 94 \%).
Experimentally \cite{bodetal91},
$<\!\sigma[^{12}{\rm{C}}(\nu_e,\nu_e^{\prime})
^{12}{\rm{C}}^*(1^+;15.1\,{\rm{MeV}})]\!>^{\rm{exp}}
=[9.5 \pm 1.8 ({\rm{stat}}) \pm 1.35 ({\rm{syst}}) ]
\times 10^{-42}\, {\rm{cm}}^2$,
where $<\!\sigma\!>$ represents the average over the
$\nu_e$ spectrum coming from the $\mu^{+}$ decay.
This agrees well with theoretical estimates
$<\!\sigma[^{12}{\rm{C}}(\nu_e,\nu_e^{\prime})
^{12}{\rm{C}}^*(1^+;15.1\,{\rm{MeV}})]\!>^{\rm{theor}}
= (9.9-10.3) \times 10^{-42} \,\rm{cm}^2$
\cite{fkk88,pm90,bp79}.
The $^{12}{\rm{C}}(\nu_e, e^-)^{12}{\rm{N}}({\rm{gnd}})$ reaction
can be identified by taking the delayed coincidence
of the recoil $e^-$ and the $e^+$ from the $^{12}$N $\beta$-decay.
The KARMEN measurement \cite{bodetal92,bodetal92b} gives
$ <\!\sigma[^{12}{\rm{C}}(\nu_e,e^{-})
^{12}{\rm{N}}({\rm{gnd}})]\!>^{\rm{exp}}
=[8.0 \pm 0.75 ({\rm{stat}}) \pm 0.75 ({\rm{syst}}) ]
\times 10^{-42}\, {\rm{cm}}^2$,
which is consistent with the earlier Los Alamos result
\cite{kraetal92}:
$ <\!\sigma[^{12}{\rm{C}}(\nu_e,e^{-})
^{12}{\rm{N}}({\rm{gnd}})]\!>^{\rm{exp}}
=[10.5 \pm 1.0 ({\rm{stat}}) \pm 1.0 ({\rm{syst}}) ]
\times 10^{-42}\, {\rm{cm}}^2$.
These experimental values are
in good agreement with recent theoretical estimates
$ <\!\sigma[^{12}{\rm{C}}(\nu_e,e^{-})
^{12}{\rm{N}}({\rm{gnd}})]\!>^{\rm{theor}}
= (8.0 -9.4 ) \times 10^{-42}\, \rm{cm}^2$
given in \cite{fkk88,don92,mp89}.

EPT calculations on the $\nu$-$^{12}$C reactions
for $E_{\nu} \gsim 100$ MeV
were carried out by several authors.
In particular, Mintz and Pourkaviani \cite{mp88}
extended the EPT calculation up to $E_{\nu}=3$ GeV.
The reliability of such extension, however,
is not well established
since the calculation involves
the nuclear form factors at very high $q^2$,
which are not known empirically except $F_M(q^2)$.
There have also been attempts to
apply EPT to nuclear systems
other than the $^{12}$C target \cite{bp79}
or to $(\nu_{\mu},\mu)$ processes \cite{mp89}.
Here again, the problem is that
there is not enough data
to determine all relevant nuclear form factors empirically.
The lack of experimental information
necessitates the introduction of
a number of additional theoretical assumptions,
which are usually motivated by
the impulse approximation results.
This {\it ad hoc} procedure largely nullifies
the predictive power of the original EPT.

Koetke \etal. \cite{koeetal92},
using the in-flight pion decay neutrino source at LAMPF,
measured the cross sections for
$^{12}{\rm{C}}(\nu_{\mu},\mu^{-})X$
and
$^{12}{\rm{C}}(\nu_{\mu},\mu^{-})
^{12}{\rm{N}}({\rm{gnd}})$
at an average neutrino energy of 202 MeV.
The cross section for the latter exclusive process is
$\sigma [\!^{12}{\rm{C}}(\nu_{\mu},\mu^{-})
^{12}{\rm{N}}({\rm{gnd}})]^{\rm{exp}}
=(1.7 \pm 0.8 \pm 0.3) \times 10^{-39}\,{\rm{cm}}^2$.
This is considerably higher than
$0.8 \times 10^{-40} \, {\rm{cm}}^2$ obtained
in the W-D method \cite{oco72,ullk72,don81}
and an ``EPT" estimate of
$0.7 \times 10^{40}\, {\rm{cm}}^2$ \cite{mp89}.
Here we have used the quotation mark
since, as discussed above,
a completely model-independent EPT analysis
is not feasible in this case.
According to Mintz and Pourkaviani \cite{mp89},
one can increase the predicted value sufficiently
by enhancing the pseudoscalar form factor $f_P$
Although it is generally believed that
the exchange-current effect quenches $f_P$
rather than enhances,
there may be other competing nuclear effects.
(For recent attempts to determine
the effective $f_P$ inside nuclei using
the radiative $\mu$-capture,
see ref.\cite{armetal92} for experiment
and ref.\cite{fw92} for  theory.)
The inclusive reaction
$^{12}{\rm{C}}(\nu_{\mu},\mu^{-})X$
will be discussed later.

\newpage

\noin
{\bf Chapter 4 \hspace{1cm}
Empirical effective operator method (EEOM)}

Going back to the ``realistic" nuclear Hamiltonian $H_N$
eq. (\ref{eq:Hnucl}), we recall that,
even if we forgo the basic problem of deriving $H_N$
from the fundamental QCD and decide to be
content with the operational usefulness of $H_N$,
it is in general prohibitively difficult
to determine true eigenstates $\Psi$'s of $H_N$
and calculate the nuclear matrix elements of
an observable ${\cal O}$ using $\Psi$'s.
This is true
even when we limit ourselves to the nucleon-only regime,
ignoring the exchange-current problem.
One therefore usually splits the space
subtended by $\Psi$'s into two parts,
a relatively manageable model space $P$
and the remainder $Q$,
and calculate observables
using wave functions $\Psi_0$'s
that belong to $P$.
The phenomenological success of the shell model
suggests choosing as $P$
the space of the lowest shell-model configurations;
if necessary, slightly excited configurations
can be included in $P$.
Once the space of $\Psi$'s is truncated
to that of $\Psi_0$'s,
$H_N$ and ${\cal O}$ must undergo
corresponding transformations,
$H_N \ra \tilde{H}_N$ and
${\cal O} \ra \tilde{{\cal O}}$
so that
$<\Psi^{\prime}|H_N|\Psi> =
<\Psi_0^{\prime}|\tilde{H}_N|\Psi_0>$
and
$<\Psi^{\prime}|{\cal O}|\Psi> =
<\Psi_0^{\prime}|\tilde{{\cal O}}|\Psi_0>$ should hold.
The nucleon-nucleon interactions
that feature in $\tilde{H}_N$
are called the {\it effective} interactions,
a somewhat historical terminology
since from today's viewpoint the original $H_N$ itself
is a highly ``effective" entity.
Similarly, $\tilde{{\cal O}}$ is called
the {\it effective} operator.
The difference between ${\cal O}$ and $\tilde{{\cal O}}$
represents the core-polarization effect.
The formal framework to obtain
the effective operators by incorporating
effects of the eliminated space $Q$
does exist (for review, see \cite{bra67,eo77}),
but applying this formalism to actual complex nuclei
encounters many practical difficulties.
Exceptionally, in the simplest cases of
closed-shell nuclei plus or minus one nucleon,
explicit calculations can be carried out
to lowest orders of perturbation,
as exemplified by the extensive work of
Tower and Khanna \cite{tk83,tow87}
and the University of Tokyo group
(Arima, Shimizu, Hyuga \etal) \cite{asbh87}.
These authors calculated the magnetic moments,
M1 transition strengths $B({\rm{M1}})$,
and the Gamow-Teller strengths $B$(GT),
taking into account
up to second-order core polarization effects.
Furthermore, the exchange-currents effects
were carefully estimated
and included in the final expressions of their
{\it effective} operators.
The results of these highly elaborate calculations
are thoroughly documented in \cite{tow87,asbh87}.

Parallel to these theoretical attempts
to {\it derive} the effective operators for the special cases,
a useful semi-empirical approach
which can cover a wider variety of nuclei
has been developed by
Wilkinson, Brown, Warburton and Wildenthal
\cite{wil73,wil74,bw83,bw85,cwb93}.
We shall refer to this approach
as the empirical effective operator method (EEOM).
As will be explained below,
EEOM finds a useful application
in estimating the cross sections
of low-energy neutrino-nucleus reactions
for nuclei up to the $sd$-shell.
We outline EEOM,
taking the Gamow-Teller (GT) operator as an example
\cite{cwb93}.

Consider a GT transition $|i> \ra |f>$
and suppose we are given reasonably realistic
nuclear wavefunctions for $|i>$ and $|f>$.
By ``realistic" we usually mean
the best available shell-model wavefunctions,
which, as today's standard demands,
are eigenvectors of
a full matrix of $\tilde{H}_N$
covering all possible configurations
within a given major shell.
The shell-model GT matrix element is given by
\ben
<GT> &\equiv&
<\!f;J_f,T_f||| \sum_{k=1}^{A}g_A
\mbox{\bm $\tau \sigma$}(k) |||i;J_i,T_i\!>, \non \\
&= & \sqrt{6}\sum_{j,j'}\, D_{j'j}
<\!j'||g_A\mbox{\bm $\sigma$}||j\!>, \lab{eq:GTbare}
\een
where the summation over $j$ and $j'$
goes over all the single-particle orbits
contained in the model space $P$,
and the one-body-transition density $D_{j'j}$
is obtained from shell-model wavefunctions
for an operator with rank in ordinary (isospin) space
of $\Delta J$ ($\Delta T$) via
\beq
D_{j'j}=\frac{<J_f,T_f|||[a^{\dagger}(j') \times
\tilde{a}(j)]^{\Delta J,\Delta T}|||J_i,T_i>}
{[(2\Delta J+1)(2\Delta T+1)]^{1/2}}.
\eeq
As stated,
even if the model space $P$ is reasonably realistic,
there are two significant sources of corrections to $<GT>$,
one from core polarization
and the other from the exchange-current effect.
If $P$ is large enough,
the core polarization effect arises from
rather highly excited states,
and consequently its dependence on $|i>$ and $|f>$
will be weak.
The same should hold for the exchange-current effect,
which comes from the eliminated hadronic degrees freedom.
In EEMO, therefore,
one simulates these effects by replacing
$<j'||g_A\mbox{\bm $\sigma$}||j>$ in eq. (\ref{eq:GTbare})
with an effective matrix element
\beq
<j'||g_A^{\rm{eff}}\mbox{\bm $\sigma$}+
     g_{LA}^{\rm{eff}} \mbox{\bm $\ell$}
  + g_{PA}^{\rm{eff}}
[Y_2 \times \mbox{\bm $\sigma$}]^{(1)}||j>,
\eeq
while keeping $D_{j'j}$ unchanged.
This replacement will give an effective GT strength
\mbox{$<GT>^{\rm{eff}}$}.
To determine the effective coupling constants,
$g_A^{\rm{eff}}$, $g_{LA}^{\rm{eff}}$
and $g_{PA}^{\rm{eff}}$
for a given major shell,
one takes a sufficiently large number of
GT decays that belong to this major shell
and whose $ft$ values,
or equivalently $B$(GT)'s,
are known experimentally.
One then adjusts
$g_A^{\rm{eff}}$, $g_{LA}^{\rm{eff}}$
and $g_{PA}^{\rm{eff}}$
in such a manner that
the $B$(GT)'s calculated with $<\!GT\!>^{\rm{eff}}$
would optimally reproduce the
chosen set of experimental $B$(GT)'s.
Once $g_A^{\rm{eff}}$, $g_{LA}^{\rm{eff}}$
and $g_{PA}^{\rm{eff}}$
are determined,
one can predict all other $B$(GT)'s
involving the same major shell.

The EEMO has proven to be highly successful
in correlating a great majority of the observed
magnetic moments, M1 strengths, and
Gamow-Teller strengths over a wide range
in the periodic table \cite{bw83,bw85,cwb93}.
Furthermore,
the effective coupling constants determined in EEOM
show reasonable agreement with
the corresponding quantities
obtained theoretically for the closed-shell plus/minus
one nucleon systems \cite{tow87,asbh87}.

Two p-shell nuclei, $^{11}$B and $^{13}$C,
have recently been considered
as useful targets to simultaneously monitor
the NC and CC reactions for the solar neutrinos.
In fact,
Raghavan, Pakvasa and Brown's proposal \cite{rpb86}
to use the $^{11}$B target is already being
intensively pursued
in the BOREX and BOREXINO projects.
We describe here
how the EEOM (or a closely related approach)
was used to estimate the cross sections
on these targets.

\vspace{0.4cm}
\noin
4.1. $^{13}$C target

The possible advantage of a $^{13}$C-enriched scintillation
counter as a solar-neutrino detector
was suggested by Arafune \etal.
\cite{afkk89} and further studied in \cite{fkkk90}.
These authors emphasized that
the replacement of $^{12}$C in a scintillator
with $^{13}$C not only makes the detector
much more sensitive to CC reactions,
but also allows efficient detection of the NC reactions,
and that
even the natural abundance of $^{13}$C ($\sim$1 \%)
may be large enough to detect NC reactions
in a large volume scintillator.
The relevant reactions are the CC reaction
\beq
^{13}{\rm{C}} + \nu_e \ra \,
^{13}{\rm{N}} + e^-, \lab{eq:NuC13CC}
\eeq
and the NC reaction
\beq
^{13}{\rm{C}} + \nu  \ra \,
^{13}{\rm{C}}^* + \nu^{\prime}. \lab{eq:NuC13NC}
\eeq
For the solar neutrino energy region.
only the following final states need to be considered \cite{ajz86}.
For the CC reactions,
$|J^{\pi}=1/2^{-},T=1/2({\rm{gnd}})>$,
$|3/2^{-},1/2(3.51\,{\rm{MeV}})>$,
$|1/2^{-},1/2(8.92\,{\rm{MeV}})>$, and
$|3/2^{-},1/2(9.48\,{\rm{MeV}})>$ in $^{13}$N,
with $Q=2.22$ MeV for the ground state transition.
For the NC reactions,
$|3/2^{-},1/2(3.68\,{\rm{MeV}})>$,
$|1/2^{-},1/2(8.86\,{\rm{MeV}})>$, and
$|3/2^{-},1/2(9.90\,{\rm{MeV}})>$ in $^{13}$C.

The total cross section $\sigma$ for the CC reaction
$\nu_e + |A> \ra e^- + |B>$ is given
in impulse approximation by
\ben
\sigma &=& \frac {G_F^2 \cos ^2\theta_c} {\pi} p_e E_e
F(Z_B,E_e)\\
  & &  \times \frac {1}{6(2J_A+1)}
 [\, |<\!B||| \mbox{\bm $\tau$} |||A\!>|^2
+ g_A^2|<\!B||| \mbox{\bm $\tau \sigma$} |||A\!>|^2\,],
\lab{eq:sigmaC13}
\een
where $g_A=-1.262$.
The Fermi matrix element contributes
only to a transition between isomultiplet members.
One often uses the reduced transition strengths
$B$(F) and $B$(GT) defined as
\ben
B({\rm{F}})  &\equiv& \sum_{M_B,M_A}
|<J_BM_B|t_{\pm}|J_AM_A>|^2
= \frac{1}{6}|<\!B||| \mbox{\bm $\tau$}|||A\!>|^2, \\
B({\rm{GT}})&\equiv&  \sum_{M_B,M_A}
|<J_BM_B|t_{\pm}\mbox{\bm $\sigma$}|J_AM_A>|^2
= \frac{1}{6}|<\!B||| \mbox{\bm $\tau \sigma$}|||A\!>|^2.
\lab{eq:BGT}
\een
The NC cross section is given by
\beq
\sigma = \frac{G_F^2}
{\pi} g_A^2 (E_{\nu}^{\prime})^2
\frac{1}{48} |<\!B||| \mbox{\bm $\tau \sigma$} |||A\!>|^2.
\eeq

For the CC,
if the $ft$ value of $\beta$ decay $|B\!> \ra |A\!>$ is known,
$\sigma$ is given model-independently in terms of $ft$:
\beq
\sigma= \frac{2\pi^2 \ln 2}{m_e^5 ft}
p_eE_e F(Z_B,E_e) \frac{2J_B+1}{2J_A+1}.
\lab{eq:sigmaft}
\eeq
This applies to
$\sigma[^{13}{\rm{C}} \ra ^{13}{\rm{N}}({\rm{gnd}})]$.
{}From
$\log(ft)^{\rm{exp}}=3.667 \pm 0.001$
for $^{13}{\rm{N}}({\rm{gnd}}) \ra
\, ^{13}{\rm{C}}({\rm{gnd}})$,
we obtain model-independently
$<\!\sigma[^{13}{\rm{C}} \ra\, ^{13}{\rm{N}}({\rm{gnd}})]\!>
= 8.12 \times 10^{-43} {\rm{cm}}^2$.
Here $<\!\sigma\!>$
stands for the cross section
averaged over the $^8$B neutrino spectrum.
(Due to the threshold energy $E_{th}=2.2$ MeV,
only the neutrinos from $^8$B decay are relevant here.)
As $^{13}{\rm{N}}({\rm{gnd}}) \ra\,
^{13}{\rm{C}}({\rm{gnd}})$
is a super-allowed transition,
one can expect a strong feeding of $^{13}$N(gnd).
For the transitions to the excited states in $^{13}$N,
for which no experimental $ft$ values are available,
one must rely on theoretical estimations.
The same applies to the NC reactions.

Fukugita \etal. \cite{fkkk90} used EEOM
to calculate these cross sections.
Thus, in eq. (\ref{eq:sigmaC13}),
$<\!B||| \mbox{\bm $\tau \sigma$}|||A\!>$
is replaced by
\beq
<\!B||| \mbox{\bm $\tau \sigma$}|||A\!>^{\rm{eff}}
\equiv
<\!B||| f_A^{\rm{eff}}\mbox{\bm $\tau \sigma$}
+f_{LA}^{\rm{eff}}\mbox{\bm $\tau \ell$}
+f_{PA}^{\rm{eff}}\mbox{\bm $\tau$}
[Y_2 \times \mbox{\bm $\sigma$}]^{(1)}|||A\!>\,.
\eeq
Correspondingly, $B$(GT) will change into
$B({\rm{GT}})^{\rm{eff}}$ [cf. eq.(\ref{eq:BGT})].
($B$(F) does not change because of charge conservation.)
For a model space $P$,
Fukugita \etal. used
the Cohen-Kurath (CK) wave functions \cite{ck65},
which are full-$p$-shell wave functions.
To determine the effective couplings,
$f_A^{\rm{eff}}$, $f_{LA}^{\rm{eff}}$
and $f_{PA}^{\rm{eff}}$,
the following three ground-state transitions were used:
$^{15}{\rm{O}}(\beta^+)^{15}{\rm{N}}$,
$^{13}{\rm{N}}(\beta^+)^{13}{\rm{C}}$, and
$^{11}{\rm{C}}(\beta^+)^{13}{\rm{B}}$.
Since $g_{LA}^{\rm{eff}} \mbox{\bm $\tau \ell$}$
is known to be very small \cite{asbh87,tow87},
$f_{LA}^{\rm{eff}}$ was fixed
at a commonly accepted value \cite{asbh87,tow87}:
$f_{LA}^{\rm{eff}} =0.011$.
The best-fit values of the remaining parameters are
$f_A^{\rm{eff}}=0.69$, $f_{PA}^{\rm{eff}}=0.19$.
These effective coupling constants
were used to calculate
$<\!B||| \mbox{\bm $\tau \sigma$} |||A\!>^{\rm{eff}}$,
$B({\rm{GT}})^{\rm{eff}}$, and the cross sections
for the reactions eqs. (\ref{eq:NuC13CC}), (\ref{eq:NuC13NC}).
The resulting $B({\rm{GT}})^{\rm{eff}}$'s were found to be
smaller than the corresponding
bare shell-model values by a factor $\sim 2$.
According to Fukugita \etal. \cite{fkkk90},
in the solar neutrino energy range,
one needs to consider
only the ground and first-excited states
in $^{13}$N for the CC reactions,
and the first excited state in $^{13}$C for the NC reaction
The calculated cross sections
for the dominant reactions leading
to the first excited states are:
$<\!\sigma[^{13}{\rm{C}} \ra \,
^{13}{\rm{N}}(3/2^{-}(3.51\,{\rm{MeV}}))]\!>
= 2.62 \times 10^{-43} {\rm{cm}}^2$
for CC, and
\mbox{$<\!\sigma[^{13}{\rm{C}} \ra\,
^{13}{\rm{C}}(3/2^{-}(3.68\,{\rm{MeV}}))]\!>
= 1.16 \times 10^{-43} \,{\rm{cm}}^2$}
for NC.

Fukugita \etal,
comparing $B({\rm{GT}})^{\rm{eff}}$ for
$^{13}{\rm{C}} \ra \,
^{13}{\rm{N}}(3/2^{-}[3.51\,{\rm{MeV}}])$
with the corresponding ``empirical" value
deduced from the ($p,n$) reaction,
assigned $\sim$30 \% uncertainty
to the calculated $\sigma$.
(The use of ($p,n$) reactions to determine
$B$(GT)'s is discussed in chapter 5.)
One could improve Fukugita \etal 's calculation
significantly by using the results of
a recent much more elaborate EEOM study \cite{cwb93}.
Chou \etal.  \cite{cwb93} have carried out
a comprehensive analysis of
Gamow-Teller $\beta$ decays
for the light ($A \leq 18$) nuclei.
The model space $P$ was taken to be
either $(0s)^4(0p)^{A-4}$ or
$(0s)^4(0p)^{A-4-n_{sd}}(1s0d)^{n_{sd}}$.
The effective GT operator was deduced for the $0p$ shell
from a least-squares fit to sixteen
$B({\rm{GT}})^{\rm{exp}}$'s,
with the results
$f_A^{\rm{eff}} \approx 0.8$, $f_{LA}^{\rm{eff}}\approx 0$,
and $f_{PA}^{\rm{eff}}\approx 0$.
Using the deduced effective operator,
eighty-three $B$(GT)'s were predicted
and compared with $B({\rm{GT}})^{\rm{exp}}$'s.
According to Chou \etal,
their EEOM analysis can reproduce
reasonably strong $B$(GT)'s belonging to the $0p$ shell
within $\sim$10 \% accuracy.
It may be worthwhile to redo the calculation of \cite{fkkk90}
taking into account Chou \etal 's latest result.
In this connection one should also
consult another detailed calculation on the p-shell nuclei
due to the Utrecht group \cite{woletal90,wol89},
which utilized the full space of
$0 \hbar \omega + 2 \hbar \omega$ configurations.

According to Fukugita \etal,
the $^{13}$C target is expected to have
$6.22 \times(1 \pm 0.1)$ SNU
from the CC reactions,
and $0.68 \times (1 \pm 0.3)$ SNU
from the NC reaction.
These correspond to
7870 CC events per kiloton yr
and 860 NC events per kiloton yr
for a $^{13}{\rm{CH}}_2$ detector;
these are both encouragingly large event rates.

The fact that number of final states that need to be taken
into account for the $^{13}$C detector is highly limited implies
the possibility of direct calibration
using neutrino beams from stopped muons.
The cross sections averaged over the $\mu^+$-decay
neutrino spectrum is estimated to be \cite{fkkk90}
\ben
<\!\sigma[^{13}{\rm{C}} (\nu_e, e^-)
^{13}{\rm{N}}(\rm{gnd})]\!>_{\mu^+ \,{\rm{decay}} }
&=& 2.22 \times 10^{-41} {\rm{cm}}^2, \non \\
<\!\sigma[^{13}{\rm{C}}(\nu_e, e^-)
^{13}{\rm{N}}(\frac{3}{2}^-;3.51 \, {\rm{MeV}})]\!>
_{\mu^+ \,{\rm{decay}} }
&=& 2.27 \times 10^{-41}{\rm{cm}}^2, \non \\
<\!\sigma[^{13}{\rm{C}}(\nu_e, \nu_e)
^{13}{\rm{C}}(\frac{3}{2}^-;3.68\, {\rm{MeV}})]\!>
_{\mu^+ \,{\rm{decay}} }
&=& 0.56 \times 10^{-41}{\rm{cm}}^2,\non \\
<\!\sigma[^{13}{\rm{C}}(\bar \nu_{\mu}, \bar \nu_{\mu})
^{13}{\rm{C}}(\frac{3}{2}^-;3.68 \, {\rm{MeV}})]\!>
_{\mu^+ \,{\rm{decay}} }
&=& 0.74 \times 10^{-41}{\rm{cm}}^2.
\een
In this estimation, the finite momentum
transfer effect is included
by multiplying $B$(F) and $B({\rm{GT}})^{\rm{eff}}$
with $j_0(qr)$.
This form factor effect reduces
$<\!\sigma\!>_{\mu^+ \, {\rm{decay}} }$ by $\sim 15 \%$,
but it is negligible ($< 1 \, \%$)
for the solar neutrino energies.
These estimates are also relevant to natural carbon
scintillator experiments,
where $^{13}$C contamination may contribute
a significant background.

\vspace{0.4cm}
\noin
4.2. \hspace{1 cm} $^{11}$B target ---
 BOREX and BOREXINO experiments
\cite{rpb86,rp88,apr90,rag91,ragetal91}

Borex is a boron-rich liquid scintillation detector
that contains approximately 200 t
of boron in the fiducial volume,
and is designed to measure the bolometric solar neutrino flux.
Borex allows the observation of four basic reactions:
\ben
\nu +^{11}{\rm{B}} &\ra & \nu' +
    [^{11}{\rm{B}}^*(E_i) \ra
\gamma(E_x) +^{11}{\rm{B}}],  \lab{eq:NuB11NC}\\
\nu_e + ^{11}{\rm{B}} &\ra & e^{-} +
[^{11}{\rm{C}}^*(E_i^{\prime})
\ra \gamma(E_x^{\prime})+^{11}{\rm{C}}],
\lab{eq:NuB11CC}\\
\nu +e^- &\ra&\nu' +e^-, \\
\bar{\nu}_e + p &\ra& n+e^+
\een
The NC process involves the three excited states
of $^{11}$B
at $E_x $= 2.12, 4.45 and 5.02 MeV,
while the CC process involves
the four states in $^{11}$C
at $E_x^{\prime}$ = 0, 2.0, 4.32 and 4.80 MeV.
The recent accumulation of solar neutrino experimental data
as discussed in chapter 1
has enhanced the importance of measuring
the solar neutrino flux coming from $^7$Be,
and the fact that the $^{11}$B target has
an appropriate detection threshold for this purpose
further increases our interest in the BOREX project.
These discussions can be found in \cite{ragetal91}.
Here we concentrate on the method used
by Raghavan, Pakvasa and Brown
in their celebrated 1986 paper \cite{rpb86}
to estimate the cross sections for the nuclear processes
eqs. (\ref{eq:NuB11NC}), (\ref{eq:NuB11CC}).
The CC reaction cross section
leading to the $^{11}$C ground state
can be directly connected to the experimental $ft$ value
for the $^{11}$C $\ra ^{11}$B $\beta$ decay
[cf. eq. (\ref{eq:sigmaft})] .
To deduce the cross sections for the other final states,
eq. (\ref{eq:sigmaC13}) without the Fermi contribution
is used.
However, the magnitudes of quenching of
the relevant $B$(GT)'s are estimated
using information on the electromagnetic process.
That is, by exploiting the facts that
the isovector part of the M1 operator also
involves \mbox{\bm $\tau \sigma$}
and that
$B({\rm{M1}})_{\rm{ISV}}^{\rm{exp}}$'s
between the ground state and the relevant excites states
in $^{11}$B are known from
$\gamma$ decay and ($e,e'$) data,
the effective $B$(GT) is determined by
assuming the proportionality
\beq
\frac{B({\rm{GT}})^{\rm{eff}}}{B({\rm{GT}})} =
\frac{B({\rm{M1}})_{\rm{ISV}}^{\rm{eff}}}
{B({\rm{M1}})_{\rm{ISV}}}.  \lab{eq:BGTM1}
\eeq
This is in fact the Walecka-Donnelly (W-D) method
\cite{wal75,dw75}
discussed in the previous chapter
rather than the EEOM as defined here.
Although the core-polarization effect
should quench $B$(GT) and $B$(M1) equally,
the exchange-current effects on $B$(GT)
can be significantly different from
those on $B$(M1).
The calculation of the exchange-current effect
\cite{tow87} indicates that the use of eq. (\ref{eq:BGTM1})
would introduce $\sim$30 \% errors.
It is therefore in general preferable to
use EEOM, in which $B({\rm{GT}})^{\rm{eff}}$
is deduced using only information
on the weak-interaction processes.
Since the effective GT operator for the p-shell
based on the elaborate EEOM is now available
\cite{cwb93},
it seems warranted to derive
new estimates of the cross sections relevant to BOREX
with the use of EEOM.
This will probably reduce the uncertainties in the cross
sections from the 30 \% to 10 \% level.
As far as the classification of calculational methods
is concerned,
we could have discussed $^{11}$B
in relation to the W-D method.
The reason why we discuss it in this chapter
is the similarity of considerations
that should go into $^{13}$C and $^{11}$B.

\newpage
\noin
{\bf Chapter 5 \hspace{0.3cm}
Determination of Gamow-Teller strengths
through $(p,n)$ reactions}

\noin
5.1 \hspace{0.4cm}Formalism

The $(p,n)$ reaction has been used extensively
to estimate the Gamow-Teller strengths $B$(GT)
that determine the cross sections of neutrino-nucleus reactions
of great astrophysical importance.
The key points here are that
the $(p,n)$ reaction, in contrast to $\beta$ decay,
is not limited by $Q$-value constraints,
and that there exists an approximate ``empirical"
proportionality between $B$(GT)
and the forward $(p,n)$ reaction cross section
at intermediate energies
\cite{gooetal80,watetal85,tadetal87}.
This proportionality was studied in great detail
by Tadeucci \etal. \cite{tadetal87}.
The essential points of their classic paper
may be summarized as follows.
Consider the $(p,n)$ reaction
\beq
p(E_p) \; + \; _ZA_N \; \ra \;
n(E_n) \; + \; _{Z+1}A_{N-1}(j),
\lab{eq:pn}
\eeq
where $j$ distinguishes various final states.
The transition leading to state $j$ may be
of the Fermi-, GT-, or mixed type;
for the sake of simplicity, however,
we first consider the pure cases
of $\alpha$-type ($\alpha$= F or GT), and describe later
how to extend the results to the mixed-type transition.
Let
$\sigma_j^{\rm{exp}}({\rm A},E_p;\, q,\omega)$
denote the observed differential cross section
with $\omega \equiv E_p-E_n$,
$\mbox{\bm $q$}$ = three-momentum transfer, and
$q=|\mbox{\bm $q$}|$.
Since the $0^{\circ}$ cross section is particularly important,
we introduce
$\sigma_j^{\rm{exp}}(0^{\circ}; \omega) \equiv
\sigma_j^{\rm{exp}}(q,\omega)|_{0^{\circ}}$,
where the momentum transfer $q$ for $0^{\circ}$ is related
to $\omega$ as $q=\omega/v_p$.
For sufficiently high incident energies and for forward angles,
the direct reaction mechanism should be dominant.
Then, for a differential cross section
$\sigma_j^{\rm{DWIA}}({\rm A},E_p;\, q,\omega)$
obtained in a DWIA calculation, we can expect
$\sigma_j^{\rm{DWIA}}(q,\omega) \approx
\sigma_j^{\rm{exp}}(q,\omega)$.
We therefore adopt as a basic ansatz the equality
\beq
\sigma_j^{\rm{exp}}(0^{\circ}; \omega)
= \sigma_j^{\rm{DWIA}}(0^{\circ}; \omega),
\lab{eq:expDWIA}
\eeq
where
$\sigma_j^{\rm{DWIA}}(0^{\circ}; \omega) \equiv
\sigma_j^{\rm{DWIA}}(q,\omega)|_{0^{\circ}}$.
In the direct-reaction picture,
a set of quantum numbers
corresponding to the isospin-flipping
and/or spin-flipping of the incident nucleon is transferred
in a {\it single} step to the target nucleus.
This single-step transfer, which could be considered as arising
from the exchange of mesons
or meson-like effective particles,
is completely analogous to
what the electroweak current does to a nucleus.
We therefore expect
$\sigma_j^{\rm{DWIA}}(q,\omega)$
for forward angles to be closely related
to the corresponding $\beta$ decay strength
$B(\alpha)$, where $\alpha=$ F or GT depending on $j$.

To guide ourselves in quantifying
this ``close" relationship,
we consider a simplified DWIA treatment
which excludes orbital angular momentum transfer
($L=0$ contribution only), and in which
the initial and final distortion effects are
taken into account using the eikonal approximation
\cite{aus70}.
This simplified treatment,
to be referred to as the eikonal DWIA calculation,
leads to an explicit proportionality relation \cite{tadetal87}:
\beq
\sigma_j^{\rm{Eik}}(q,\omega)
=K(E_p,\omega)\, \exp[-xA^{1/3}+p(\omega)]\,
|J_{\alpha}|^2\, \exp(-\frac{1}{3}q^2<r_{\alpha}^2>)
B(\alpha)_j. \lab{eq:Tad2.17}
\eeq
Although the actual deduction of $B(\alpha)_j$ always uses
the $0^{\circ}$ cross sections,
we are considering here
general forward-angle scatterings,
in which $q$ and $\omega$ are independent variables.
In the present context it is useful
to treat these two variables separately,
since the $q$- and
$\omega$-dependence of the cross sections
arises from quite different physical origins.
The first factor in eq. (\ref{eq:Tad2.17}) is kinematic in origin,
and its explicit form is
\beq
K(E_p,\omega)= \frac{E_pE_n}{\pi^2}\, \frac{k_n}{k_p}.
\lab{eq:Tad2.18}
\eeq
The second factor represents the effect of distortion
of the initial and final nucleon waves,
parametrized in terms of the square-well optical potentials
$U_{i,f} = -(V_{i,f}+ {\rm{i}}W_{i,f})$
with radius $R=r_0 A^{1/3}$;
$x=4W_ir_0/v_p$ and
$p(\omega)$ arises from the difference between the
initial and final optical potentials.
One could calculate $p(\omega)$ in the eikonal DWIA,
but it will prove more useful to determine it empirically.
$J_{\alpha}$ is,
apart from a correction for the antisymmetrization of
the valence nucleon and target nucleus,
the volume integral of the N-N $t$-matrix
in a channel corresponding to $\alpha$:
\beq
J_{\alpha} = \int V_{\alpha}(r)\,d^3r\,\, +\,\,
({\rm{Pauli\!-\!exchange \,\, term}}), \lab{eq:VolumeInt}
\eeq
where $V_{\alpha}(r)$ is the radial dependence
of the relevant $t$-matrix;
$t(1,2) = V_{\rm{F}}(r_{12})\,\vec{\tau}_1 \cdot \vec{\tau}_2$
or
$t(1,2) = V_{\rm{GT}}(r_{12})\,
\vec{\tau}_1 \cdot \vec{\tau}_2\,\,
\vec{\sigma}_1 \cdot \vec{\sigma}_2$.
The fourth factor in eq. (\ref{eq:Tad2.17})
accounts for the momentum-transfer dependence
coming from the overall nuclear size effect
as well as the $q$-dependence in the $L=0$ Bessel transform
of $V_{\alpha}(r)$.
We factor out the $q$- and $\omega$-dependence
in eq. (\ref{eq:Tad2.17}) as
\beq
\sigma_j^{\rm{Eik}}(q,\omega)
=F^{\rm{Eik}}(q,\omega) \,
\sigma_j^{\rm{Eik}}(q=0,\omega=0),\lab{eq:Tad1.1a}
\eeq
where
\ben
F^{\rm{Eik}}(q,\omega) &\equiv&
\frac{\sigma_j^{\rm{Eik}}(q,\omega)}
        {\sigma_j^{\rm{Eik}}(q=0,\omega=0)}, \non\\
&=& \frac{K(E_p,\omega)}{K(E_p,\omega=0)}
\,\exp(-\frac{1}{3}q^2<r^2>)
\, \exp[p(\omega)-p(\omega=0)]\,. \lab{eq:Tad2.21}
\een
Similarly, we define
\beq
F_j^{\rm{DWIA}}(q,\omega) \equiv
\frac{\sigma_j^{\rm{DWIA}}(q,\omega)}
        {\sigma_j^{\rm{DWIA}}(q=0,\omega=0)}.
\lab{eq:FDWIA}
\eeq
By definition,
$F^{\rm{Eik}}(q=0,\omega=0)
= F_j^{\rm{DWIA}}(q=0,\omega=0)=1$.
The separability of $B(\alpha)_j$
in $\sigma_j^{\rm{Eik}}(q,\omega)$
[eq. (\ref{eq:Tad2.17})] implies that
\beq
\hat{\sigma}_j^{\rm{Eik}}(A, E_p)
\equiv \sigma_j^{\rm{Eik}}(A, E_p;\, q=0,\omega=0)/
B(\alpha)_j, \lab{eq:Tad1.1b}
\eeq
is independent of individual transitions $j$,
except for the distinction between the F- and GT-types.
We therefore write
$\hat{\sigma}_{\alpha_j}^{\rm{Eik}}(A, E_p)$
instead of $\hat{\sigma}_j^{\rm{Eik}}(A, E_p)$.
Its explicit form is
\beq
\hat{\sigma}_{\alpha_j}^{\rm{Eik}}(A, E_p)
=K(E_p, \omega=0)\,
\exp[-xA^{1/3}+p(\omega=0)]\,|J_{\alpha_j}|^2.
\lab{eq:Tad1.1bbb}
\eeq
We also introduce
\beq
\hat{\sigma}_j^{\rm{DWIA}}(A, E_p)
\equiv \sigma_j^{\rm{DWIA}}(A, E_p;\, q=0,\omega=0)/
B(\alpha)_j \, . \lab{eq:Tad1.1c}
\eeq
The hatted quantities,
$\hat{\sigma}_{\alpha_j}^{\rm{Eik}}(A, E_p)$ and
$\hat{\sigma}_j^{\rm{DWIA}}(A, E_p)$,
which represent $(p,n)$ cross sections
one would obtain for $q=0$, $\omega=0$
if $B_{\alpha_j}=1$,
are called the {\it unit cross sections}.
In terms of the unit cross section
$\hat{\sigma}_{\alpha_j}^{\rm{Eik}}(A, E_p)$,
the separability encoded in eq. (\ref{eq:Tad2.17})
is expressed as
\beq
\sigma_j^{\rm{Eik}}(q,\omega) =
F^{\rm{Eik}}(q,\omega)\,
\hat{\sigma}_{\alpha_j}^{\rm{Eik}}(A, E_p)
\,B(\alpha)_j.
\lab{eq:Tad1.1Eik}
\eeq
Note that, since the way  $B(\alpha)_j$ enters into
$\hat{\sigma}_j^{\rm{DWIA}}(A, E_p)$
is yet to be determined,
the definition in eq. (\ref{eq:Tad1.1c})
is nothing more than a convenient numerical normalization;
no explicit expression such as
eq. (\ref{eq:Tad1.1bbb}) can be derived therefrom..

Guided by eq. (\ref{eq:Tad1.1Eik}),
we try to derive an approximate proportionality
between $B(\alpha)_j$ and
$\sigma_j^{\rm{exp}}(A, E_p;\, q,\omega)$
or, more specifically,
between $B(\alpha)_j$ and
$\sigma_j^{\rm{exp}}(A, E_p;\, 0^{\circ}; \omega)$.
The first step is to check to what extent the
eikonal DWIA can reproduce the numerical results of
the full DWIA calculation.
For the purpose of comparing these two approximations
concerning the reaction mechanism,
one can use simplified nuclear wave functions.
Tadeucci \etal. \cite{tadetal87},
employing extreme single-particle wavefunctions,
compared the eikonal DWIA and the full DWIA
over a wide range of $A$ and $E_p$.
Their results indicate
that, if the parameters appearing in
$F^{\rm{Eik}}(q,\omega)$ [eq. (\ref{eq:Tad2.21})]
are treated as adjustable parameters,
$F^{\rm{Eik}}(q,\omega)$ can fit
$F^{\rm{DWIA}}(q,\omega)$ reasonably well.
The {\it effective} $F$ factor obtained
through this fit is denoted by $F(q,\omega)$.
Thus
\beq
F(q,\omega) \equiv
\frac{K(E_p,\omega)}{K(E_p,\omega=0)}
\,\exp(-\frac{1}{3}q^2<\tilde{r}^2>)
\, \exp[\tilde{p}(\omega)-\tilde{p}(\omega=0)]\,.
\lab{eq:Tad2.21eff}
\eeq
where $\tilde{r}$ is the optimal radial parameter,
while $\tilde{p}(\omega)$ is the optimal functional form of
$p(\omega)$.
(Quadratic functions suffice to fit the DWIA results.)
Furthermore, by fixing the parameters
in eq. (\ref{eq:Tad1.1bbb})] to best-fit
$\hat{\sigma}_j^{\rm{DWIA}}(A, E_p)$,
we can determine
the optimized $\hat{\sigma}_{\alpha_j}^{\rm{Eik}}(A, E_p)$,
which we denote by
$\hat{\sigma}_{\alpha_j}(A, E_p)$.
According to Tadeucci \etal. \cite{tadetal87},
$\hat{\sigma}_{\alpha_j}(A, E_p)$
is a smooth function of $A$ and $E_p$ and
is able to reproduce the average behavior of
$\hat{\sigma}_j^{\rm{DWIA}}(A, E_p)$ satisfactorily;
for GT transitions ($\alpha_j=$ GT),
the scatter around $\hat{\sigma}_{\alpha_j}(A, E_p)$
due to the $j$-dependence is about 7 \%.
Using these results as well as
eqs. (\ref{eq:FDWIA}), (\ref{eq:Tad1.1c}),
we obtain
\ben
\sigma_j^{\rm{DWIA}}(q,\omega) &=&
F_j^{\rm{DWIA}}(q,\omega)
\hat{\sigma}_j^{\rm{DWIA}}(A, E_p) \,B(\alpha)_j, \non \\
& \cong & F(q,\omega)\,\hat{\sigma}_{\alpha_j}(A, E_p)
\,B(\alpha)_j\,. \lab{eq:Tad1.1eff}
\een
Since the various assumptions
made in deriving this result are best justified
for $0^{\circ}$ scattering,
we specialize ourselves here to the $0^{\circ}$ case.
Then, invoking the basic assumption
eq. (\ref{eq:expDWIA}), we arrive at
\beq
\sigma_j^{\rm{exp}}(0^{\circ}; \omega)
\cong
F(q,\omega)|_{0^{\circ}}
\,\hat{\sigma}_{\alpha_j}(A, E_p) \,B(\alpha)_j\,.
\lab{eq:final}
\eeq
Note that once the type of a transition
($\alpha_j$ = F or GT) is specified,
the factor
$F(q,\omega)|_{0^{\circ}}
\,\hat{\sigma}_{\alpha_j}(A, E_p)$
is independent of individual transitions specified by $j$,
apart from the easily calculable kinematical dependence
through $\omega$.
Thus eq. (\ref{eq:final}) represents
an approximate proportionality between
$\sigma_j^{\rm{exp}}(0^{\circ}; \omega)$
and $B(\alpha)_j$.
The extension of this relation to the case of
a mixed transition is
\beq
\sigma_j^{\rm{exp}}(0^{\circ}; \omega)
\cong
F(q,\omega)|_{0^{\circ}} \times
\,[\hat{\sigma}_{\rm{F}}(A, E_p) \,B({\rm{F}})_j\,
+\, \hat{\sigma}_{\rm{GT}}(A, E_p) \,B({\rm{GT}})_j].
\lab{eq:mixed}
\eeq
As a matter of fact, $F(q,\omega)$ slightly
varies depending on $\alpha_j$= F or GT,
through the $\alpha_j$-dependence of
$\tilde{r}$ in eq. (\ref{eq:Tad2.21eff}).
This difference, however, is negligible in practice,
and was ignored in the above expression.

In the literature one often encounters another expression
which is equivalent to eq. (\ref{eq:final}).
Dropping the optical potential
in what is called here the eikonal DWIA,
we consider the plane-wave impulse approximation
(PWIA).
The unit cross section in PWIA,
$\hat{\sigma}_j^{\rm{PWIA}}(A,E_p)$,
is given simply by ignoring the exponential factor
in eq. (\ref{eq:Tad1.1bbb}):
\beq
\hat{\sigma}_j^{\rm{PWIA}}(A,E_p)=
K(E_p,\omega=0) |J_{\alpha_j}|^2\,.\lab{eq:SigPWEik}
\eeq
The ratio
\beq
N^D  \equiv
\frac {\hat{\sigma}_{\alpha_j}^{\rm{Eik}}(A, E_p)}
{\hat{\sigma}_j^{\rm{PWIA}}(A,E_p)}
=\exp[-xA^{1/3}+p(\omega=0)] , \lab{eq:ND}
\eeq
represents the optical potential effect for
the $0^{\circ}$ scattering, and is called
the distortion factor.
Eqs. (\ref{eq:Tad1.1b}),
(\ref{eq:Tad1.1bbb}) and (\ref{eq:ND}) lead to
\beq
\sigma_j^{\rm{Eik}}(A, E_p;\, q=0,\omega=0)=
K(E_p, \omega=0)\,N^D\, |J_{\alpha_j}|^2\,
B(\alpha)_j\, .  \lab{eq:NDform}
\eeq
This is completely equivalent to the $0^{\circ}$
value of eq. (\ref{eq:Tad1.1Eik}).
Furthermore, upon optimizing the parameters appearing
in eq. (\ref{eq:NDform}) to fit the full DWIA results,
eq. (\ref{eq:NDform}) would become
equivalent to eq. (\ref{eq:final}).

The important question is
how reliable eq. (\ref{eq:final}) is.
As mentioned above,
there is an intrinsic uncertainty ($\sim$ 7 \%)
due to the scatter of
$\sigma_j^{\rm{DWIA}}(q,\omega)$
around the smooth function
$\hat{\sigma}_{\alpha_j}(A, E_p)$.
In fact, an even larger ambiguity arises
from the optical-model dependence of
$\sigma_j^{\rm{DWIA}}(q,\omega)$
and $\hat{\sigma}_{\alpha_j}(A, E_p)$.
These unit cross sections can vary
up to $\sim$ 30 \%
within reasonable choices of the optical potential.
(The ratio $\sigma_j^{\rm{DWIA}}(q,\omega)/
\hat{\sigma}_{\alpha_j}(A, E_p)$ is not sensitive to
this change.)
Thus, at present, the proportionality coefficient
in eq. (\ref{eq:final}) cannot be calculated
with accuracy
better than $\sim$ 30 \%.

A useful alternative is to determine
the proportionality coefficient empirically
using transitions for which both
$\sigma_j^{\rm{exp}}(0^{\circ}; \omega)$
and $B({\alpha})_j^{\rm{exp}}$ are known.
For each of these transitions
one can deduce the {\it empirical} unit cross section via
\beq
[\hat{\sigma}_{\alpha}^{\rm{emp}}]_j \equiv
\frac{\sigma_j^{\rm{exp}}(0^{\circ}; \omega)}
{F(q,\omega)|_{0^{\circ}}\,B({\alpha})_j^{\rm{exp}}}.
\lab{eq:SigHatEmp}
\eeq
If the proportionality relation eq. (\ref{eq:final})
is a good description of reality,
$[\hat{\sigma}_{\alpha}^{\rm{emp}}]_j$'s
should lie close to the smooth curve corresponding to
$\hat{\sigma}_{\alpha_j}(A, E_p)$.
For the GT process,
Tadeucci \etal. \cite{tadetal87} studied
a large number of transitions
covering from $A=6$ to $A=162$
for various incident energies $E_p$.
Tadeucci \etal. refer to the proportionality
among transitions with a common pair of initial and final
nuclides as specific proportionality.
The proportionality within transitions involving
different pairs of nuclides is called general proportionality.
Tadeucci \etal.'s results indicate that
the specific proportionality holds
typically to an accuracy of $\sim 5$\%.
Thus the use of empirical data
helps to reduce ambiguities which at present cannot
be controlled theoretically.
On the other hand, the same authors report that
the general proportionality
may be uncertain by as much as $\sim$ 50 \%.
In particular, the cases involving odd-$A$ targets
exhibit the largest deviations
from the predicted smooth curve.
As far as the general proportionality is concerned,
therefore,
the model consideration described here is open
to further improvements.
The even-odd mass-number effect in the optical potential
was cited as a possible solution for this problem
\cite{tadetal87}.

The above argument suggests
as a practical way to use the ``proportionality" the following.
To infer $B({\alpha})_j$ pertaining to a transition in a
mass-$A$ nucleus, find a transition $j'$ for
which $B({\alpha})_{j'}^{\rm{exp}}$ is known
from a $\beta$-decay experiment
and which belongs either to the same mass-$A$ nucleus
or to the closest possible neighboring nucleus.
Measure $\sigma_j^{\rm{exp}}(0^{\circ}; \omega)$
and $\sigma_{j^{\prime}}^{\rm{exp}}(0^{\circ}; \omega)$.
Compute $[\hat{\sigma}_{\alpha}^{\rm{emp}}]_{j^{\prime}}$
from eq. (\ref{eq:SigHatEmp}).
Obtain $B({\alpha})_j$ using eq. (\ref{eq:final})
together with the assumption
$\hat{\sigma}_{\alpha_j}
=[\hat{\sigma}_{\alpha}^{\rm{emp}}]_{j^{\prime}}$.
This is in fact essentially the method
that was used by Goodman \etal.
in their seminal paper \cite{gooetal80},
prior to the development of the detailed
treatment described in \cite{tadetal87}.

As a particular case of the specific proportionality
Tadeucci \etal. also analyzed mixed-type transitions
[eq. (\ref{eq:mixed})]
leading to the analog states of target nuclei.
A very useful empirical rule to relate the $B$(GT)
of an analog transition to the $B$(F) in the same transition
was reported in \cite{tadetal87}.  Note that $B$(F)
is given model-independently by $B$(F) = $N-Z$.

We add that all these results can also be stated
using eq. (\ref{eq:NDform}),
which is equivalent to eq. (\ref{eq:final}).

\noin
5.2 \hspace{0.4cm} Experimental studies

Systematic experimental studies of the GT-strength in nuclei
with the use of intermediate-energy $(p,n)$ reactions
dramatically started with two seminal articles in 1980
\cite{baietal80,gooetal80}.
The first paper \cite{baietal80}
clearly demonstrated the existence
of the giant GT resonance
predicted by Ikeda, Fujii and Fujita as early as 1963
\cite{iff63}.
In the other paper \cite{gooetal80},
Goodman \etal. \cite{gooetal80}
established the approximate proportionality
eq. (\ref{eq:final}) between $B$(GT)'s
and the forward $(p,n)$ reaction cross section
and emphasized that
this proportionality could be a powerful tool
to extract GT strengths for transitions inaccessible to
$\beta$ decay.
Indeed this tool has been used extensively in
many subsequent works
\cite{watetal85,tadetal87,andetal85,andetal87,andetal91}.
Among the main objectives of these investigations are:
(i) Determining GT-strength distributions
over wide excitation-energy ranges
and comparing them with shell-model predictions
as well as with the GT-strength sum rule.
(see e.g. \cite{bw88,andetal91});
(ii) Extracting $B$(GT)'s for particular transitions
which play important roles in astrophysical
neutrino detections.
Although items (i) and (ii) are correlated,
the latter is of more direct relevance to our discussion,
and we mostly concentrate here on (ii).

A list of neutrino-nucleus reaction cross sections
that are relevant to solar neutrino detection
and that have been determined
with the help of $(p,n)$ reaction data includes
$^{19}$F, $^{37}$Cl \cite{rapetal81},
$^{40}$Ar, $^{71}$Ga \cite{kroetal85},
$^{81}$Br \cite{kroetal87,sug93},
$^{98}$Mo \cite{rapetal85},
$^{115}$In, $^{127}$I \cite{sug93}, $^{205}$Tl \cite{krof87}.
A thorough description of specific
solar neutrino experiments
that use these nuclear targets
and a critical discussion of uncertainties
in the estimates of the relevant cross sections
can be found in Bahcall and Ulrich \cite{bu88},
and in Bahcall \cite{bah89}.
There is no need
to repeat what is discussed extensively
in these references
but, in view of great current interest in the results
of GALLEX \cite{GALLEX} and SAGE \cite{SAGE1,SAGE2},
it may be useful to remind the reader that
the standard solar model value of
$132^{+20}_{-17}$ SNU for a $^{71}$Ga detector,
given in \cite{bu88,bah89} and used in the analyses
of the GALLEX and SAGE data,
already includes rather generous error estimates
of the $B$(GT)'s for $^{71}$Ge excited states.
[The cross section for the ground-state transition
can be determined from
the $ft^{\rm{exp}}(^{71}{\rm{Ge}}\ra \, ^{71}{\rm{Ga}})$
without using the $(p,n)$ reaction,
and hence has little uncertainty.]

We mention here two post-Bahcall-Ulrich developments.
One is concerned with the
\beq
\nu_{\rm{e}} \, + \, ^{127}{\rm{I}} \, \ra \,
{\rm e}^- \, + \, ^{127}{\rm{Xe}}\,\,
(Q \, =\, 0.789\,{\rm{MeV}}) \, .  \lab{eq:Iodine}
\eeq
Originally, Haxton \cite{hax88} advocated
the usefulness of an iodine detector
for testing whether the solar neutrino flux is constant
as a function of time.
Our interest in the iodine detector
has recently received an additional boost
from the fact that it is sensitive to
the $^{7}$Be neutrinos as well as to the $^{8}$Be neutrinos.
The results of the GALLEX and SAGE experiments
\cite{GALLEX,SAGE1,SAGE2} have further enhanced
the importance of examining the energy dependence
of solar neutrino deficits with the use
of nuclear targets with different reaction $Q$-values.
The $^{127}$I target is a promising candidate
to cover the region between the $^{71}$Ga experiment
and the water-Cerenkov-counter and $^{37}$Cl experiments.
The main issue with $^{127}$I experiments is
how to estimate the cross section of eq. (\ref{eq:Iodine})
reliably, and this problem is currently studied
with great intensity.
\cite{workshop93}.
For an update of $(p,n)$-reaction determination
of the $B$(GT)'s for $^{127}$I
as well as for$^{71}$Ga, $^{81}$Br,
see e.g. Sugarbaker \cite{sug93}.
For the current status of calibrating
the $^{127}{\rm{I}}(\nu_e,e^-)\,^{127}{\rm{Xe}}$
cross section with the LAMPF muon-decay neutrinos,
see e.g. Wildenhain \cite{wil93};
for the nuclear structure consideration of the relevant
$B$(GT)'s, see Engel \etal. \cite{epv91}
and Haxton \cite{hax93}.

Another topic that has recently been attracting
much attention will be discussed in the next section.

\vspace{0.5cm}
\noin
5.3 \hspace{0.4cm} The low-energy
$\nu_e+ \,\Cl37 \ra e^- + \A37 $ reaction

As is well known, the reaction
\beq
\nu_e+ \,\Cl37 \ra e^- + \A37 , \lab{eq:Cl37nu}
\eeq
was the first to be used for detecting the solar neutrinos.
It is also well known that the determination of
the relevant cross sections presents a unique case.
That is, there exists $\beta^+$ decay
\beq
^{37}_{20}\rm{Ca}_{17} \ra
^{37}_{19}\!\rm{K}_{18} + e^+ + {\nu}_e, \lab{eq:Ca37}
\eeq
which is the mirror process of
\beq
^{37}_{17}\rm{Cl}_{20} \ra
^{37}_{18}\!\rm{Ar}_{19} + e^- + \bar{\nu}_e,
\lab{eq:Cl37beta}
\eeq
and all of the $^{37}$Ar states that are relevant
to the solar neutrino reaction eq. (\ref{eq:Cl37nu})
have their counterparts in the final $^{37}$K states
of the $^{37}$Ca $\beta^+$-decay eq. (\ref{eq:Ca37}),
since this decay has a very large energy release
($Q_{\rm{EC}}$=11.64 MeV).
Therefore, using the experimental $ft$-values
for individual final states in eq. (\ref{eq:Ca37}),
one can determine model-independently
low-energy cross sections for eq. (\ref{eq:Cl37nu})
[{\it cf}. eq. (\ref{eq:sigmaft})].
The mirror-asymmetry problem \cite{wil78}
can in principle affect this procedure to some extent.
The mirror asymmetry parameter $\delta$ is
defined by
$\delta \equiv (ft)^+/(ft^-) \, - \, 1$ for a pair
of mirror $\beta^+$ and $\beta^-$ decays.
With no charge-asymmetric effects
in nuclear structure, and
in the absence of second-class currents (SCC)
in the weak-interaction current,
$\delta$ should be identically zero.
Experimentally, $\delta^{\rm{exp}} \lsim 0.1$
for even-$A$ systems, and
$\delta^{\rm{exp}}$ can be as large as $\sim$0.2
for odd-$A$ nuclei \cite{wil78}.
(It is commonly accepted that the finite values
of $\delta^{\rm{exp}}$ are caused by the ``trivial" nuclear
charge-asymmetry effects rather than by the SCC;
no explicit calculations
that can explain the large $\delta^{\rm{exp}}$
in the odd-A cases have ever been presented, however
\cite{wil78,ob89}.)
The $\delta$ effect should in principle exist
in the $A=37$ case as well.
Fortunately, however, the solar-neutrino
reaction on $^{37}$Cl is dominated by the
feeding of the analog-state ($E_x$= 4.98 MeV)
whose strength is primarily governed
by the well-known $B({\rm{F}})=N-Z$,
a situation that renders the estimation of
$\sigma[^{37}{\rm{Cl}}(\nu_e,e^-)\,^{37}{\rm{Ar}}]$
significantly less sensitive to the mirror-asymmetry problem
than the general magnitude of $\delta^{\rm{exp}}$
may suggest.
As one goes beyond the solar neutrino energies,
however, the contributions of the higher-lying states
become progressively more important and hence
the $\delta$ problem may acquire a practical significance.

Another problem whose importance increases
as $E_{\nu}$ becomes larger than
the solar neutrino energies is
the discrepancy between the $B$(GT)'s deduced
from the $^{37}$Ca $\beta$ decay \cite{sgc74}
and those obtained from the $(p,n)$ reaction
\cite{rapetal81}.
In discussing this problem, it is useful
to consult the shell-model prediction \cite{bw83}
obtained with the use of
the empirical transition operator method (cf. chapter 4)
based on the Wildenthal wavefunctions \cite{wil84}.
Now, while
$^{37}{\rm{Ca}} \ra \, ^{37}{\rm{K}} + \beta^+$ data
\cite{sgc74} give $B$(GT)'s in fair agreement
with the shell-model values
for $E_x \leq 5.7$ MeV, they yield too small $B$(GT)'s
for the levels in the $E_x\,= \,6.5-8.5$ MeV range.
According to Adelberger and Haxton \cite{ah87},
this discrepancy arises because
Sextro \etal. \cite{sgc74} assumed
that the delayed proton emission
from $^{37}$K leads only to the $0^+$ ground state
of $^{37}$Ar,
thus ignoring the expected substantial feeding of the
$2^+(3.827 \, \rm{MeV})$ state;
if the decay to this $2^+$ state is properly taken into account,
the resulting $B$(GT)'s are rather close to the shell-model
prediction.
The $(p,n)$ data \cite{rapetal81}
indicate a strong (summed) $B$(GT)
in the $E_x\,= \,6.5-8.5$ MeV region
in agreement with the shell-model prediction.
The overall agreement between the $B$(GT)'s obtained
from the $(p,n)$ reaction \cite{rapetal81}
and the shell-model values \cite{bw83} is reasonable
except for levels in the vicinity of the analog state
($E_x=4.98$ MeV),
for which the $\Cl37(p,n)\A37$ reaction data
give too small $B$(GT)'s.
Adelberger and Haxton \cite{ah87}
suggested that this depletion of the GT strengths was caused by
an incorrect subtraction procedure of the strong Fermi
contribution.

Garc\'ia \etal. \cite{garetal91}
recently carried out a high-resolution, low-background experiment
on $^{37}$Ca $\beta^+$ decay, drastically improving
the time-honored work of Sextro, Gough and Cerny \cite{sgc74}.
For $E_x \leq 5.5$ MeV,
the $B$(GT)'s obtained by Garc\'ia
are in fair agreement with those of \cite{sgc74},
except that Garc\'ia \etal. see no evidence for feeding of
a 4.679 Mev level in $^{37}$K.
On the other hand, Garc\'ia \etal. find much more strength
at $E_x \geq$ 5.5 MeV, and the integrated $B$(GT)
is roughly twice that of Sextro \etal.
Furthermore, Garc\'ia \etal. identify $\approx$50 \% more
GT strength below $E_x$=8.0 MeV than was inferred
from the $(p,n)$ data \cite{rapetal81}.
According to Adelberger \etal. \cite{adeetal91},
the integrated $B$(GT)$^{\rm{exp}}$
up to $E_X$=8 MeV does not require
the renormalization of $g_A$ (see also \cite{andetal91}),
and the quenching of $g_A$ indicated by
shell-model analyses \cite{bw85,bw88}
of the $\beta$ decay data may in fact suggest
that the shell-model calculations \cite{bw85,bw88}
tend to shift the $B$(GT)'s to the lower excited levels.
Since the $\beta$-decay data concerns only low-lying states,
the shell-model fit of the observed $\beta$-decay strengths
leads to the apparent quenching of $g_A$.

Critical examinations of
various aspects of Garc\'ia \etal.'s analyses
\cite{garetal91,adeetal91}
have been made in \cite{rs92,gooetal92,abrg92}.
The comparison
of the latest $^{37}$Ca $\beta^+$ decay result
with the ($p,n$) data points out
among other things the following two problems:
(i) On the $(p,n)$ side, the determination of the
$B$(GT) in the isobaric-analog state region
($E_x \approx$ 5 MeV)
is quite delicate because of the dominant Fermi transition
and the limited experimental energy resolution
\cite{rs92};
(ii) On the $\beta$-decay side, the assumption used in
\cite{garetal91} that all the $^{37}$K daughter states
above the proton threshold should decay by proton emission
is not tenable \cite{gooetal92,abrg92}.
In fact, the 3.24 MeV level in $^{37}$K
has $\Gamma_{\gamma}/\Gamma_p \approx 40$
\cite{ilietal93}, and this can change significantly
Garc\'ia \etal.'s analysis of the delayed-proton emission.

A detailed comparison between the updated values of
$B({\rm{GT}})^{\rm{exp}}$
and the improved shell-model prediction
was made by Brown \cite{bro92}.
The results of a recent re-measurement
\cite{wel92} of the $^{37}{\rm{Cl}}(p,n)^{37}{\rm{Ar}}$
are expected to further clarify the issue.

It is to be remarked that,
although from a nuclear-structure point of view
it is important and interesting to
settle the existing quantitative difference
between the $^{37}$Ca $\beta^+$ decay result
and the $(p,n)$ data,
the influence of this discrepancy
on the solar neutrino cross section is rather modest;
for instance, the $B$(GT)'s
deduced by Garc\'ia \etal. \cite{garetal91}
implies that
$\sigma[^{37}{\rm{Cl}}(\nu_e,e^-)^{37}{\rm{Ar}}]$
for the $^8$B neutrinos would be enhanced
by 6\% over the literature value.

\newpage
\noin
{\bf Chapter 6 \hspace{1cm}
Inclusive neutrino-nucleus reactions
- Fermi gas model}

\noin
6.1\hspace{0.3cm} Overview

Our interest in intermediate- and high-energy
neutrino-nucleus reactions was greatly enhanced
by an intriguing issue raised by Davis
\cite{dav86}.
According to Davis,
the $^{37}\rm{Ar}$ production rate in the Cl experiment
has a significant time structure
which may be correlated with the occurrence of
large solar flares \cite{dav86}.
A possible explanation is that,
when protons accelerated by the flare activity
collide with the solar atmosphere,
they may produce
a sufficient fluence of
medium- and high-energy neutrinos
(to be called the solar-flare neutrinos)
to cause significant excess counts of $\A37$.
The typical energies of the solar-flare neutrinos are
$E_{\nu} \approx$ a few hundred MeV,
and their high energy tail can
even extend to the GeV region.
The original suggestion \cite{dav86}
was based on a crude estimate
of the $\nu$-$^{37}$Cl cross section at $\Enu=$100 MeV,
but quantitative considerations would require
more elaborate estimates.
Davis's argument based on the Homestake experiment
was followed by the suggestion
\cite{fuk87,bah88,hiretal88b,fkkk89}
that large water \v{C}erenkov detectors
such as the Kamiokande and IMB facilities
can furnish a powerful tool
to measure the fluence of solar-flare neutrinos.
The point is that because of the high energies
of the flare-neutrinos the $^{16}\rm{O}$ nuclei
that are abundantly contained in a water detector
become extremely efficient targets for detecting
neutrinos via inclusive reactions \cite{hax87}.
A similar situation exists for
the $^{12}\rm{C}$ nuclei in scintillation counters \cite{bah88}.
These arguments call for reasonably reliable estimates
of neutrino-nucleus reaction cross sections
for $\Enu$ typically up to a few hundred MeV.
Two different problems are involved here.
For radio-chemical experiments that count
only those events that lead to particle-bound states,
we must calculate {\it semi-inclusive} cross sections,
whereas water \v{C}erenkov or liquid scintillation
experiments involve the usual inclusive reaction cross sections.
Relegating the discussion of the semi-inclusive reactions
to chapter 7, we survey here recent works
concerning the inclusive reactions.

The inclusive neutrino-nucleus reactions
are also important in relation to the atmospheric neutrinos.
When primary cosmic rays strike the atmosphere,
they produce pions and kaons that subsequently
decay into muons and muon-neutrinos,
and much less frequently electrons and electron-neutrinos.
The muons further decay into electron-neutrinos
and muon-neutrinos.  The neutrinos of this origin,
called the atmospheric neutrinos,
have been observed in the range of
a few hundred MeV to several GeV with
the large water \v{C}erenkov detectors
\cite{hiretal88a,bioetal88}.
The interactions of the atmospheric neutrinos
in water are important for two main reasons.
One is that these interactions
constitute the principal background
for nucleon decay experiments and
extraterrestrial neutrino searches.
The second reason is an important current issue
concerning the atmospheric $\nu_{\mu}/\nu_e$ ratio.
The above-described origin of the atmospheric neutrinos
implies that there should be roughly two muon-neutrinos
for each electron-neutrino.
(Here ``neutrinos" mean both neutrinos and
antineutrinos.)
The experimental data \cite{hiretal88a,bioetal88},
however, indicates that the muon-neutrinos are
significantly less abundant than expected.
The cross sections for
$^{16}{\rm{O}}(\nu_{\ell},\ell^-)X({\rm{anything}})$
and
$^{16}{\rm{O}}(\bar \nu_{\ell},\ell^+)X({\rm{anything}})$
feature importantly in analyzing these data.

\vspace*{0.4cm}
\noin
6.2 \hspace{0.3cm} Calculational methods

The study of inclusive nuclear responses
to medium- or high-energy electroweak probes
has a vast accumulation of literature
\cite{defw66,wal75,dw75,bf64,bls71,sm72,bugetal79,go86}.
The closure approximation \cite{bf64,bls71}
and the quasi-free scattering approach
\cite{wal75,sm72,bugetal79,go86}
are the most commonly used methods.
Kuramoto \etal. \cite{fkkk89,kfkk90}
have recently made a detailed comparison of
these two formalisms
and recommend that the closure approximation
not be used for quantitative estimations.
We therefore concentrate here
on the quasi-free scattering approach.

We are concerned with the inclusive reactions
\ben
     \nu_{\ell}\,+ \rm{A}
&\ra& \ell^- \, + \, X(\rm{anything}),
 \lab{eq:nuAX}\\
\bar \nu_{\ell}\ + \rm{A}
&\ra& \ell^+ \, + \, X(\rm{anything}).
\lab{eq:nubarAX}
\een
In the quasi-free scattering (QFS) approach
\cite{defw66,wal75}
the transition amplitudes for the above processes
are obtained by summing the transition amplitudes
for neutrino-nucleon scattering
over $A$ nucleons characterized
by the initial and final nuclear wavefunctions.
In the so-called {\it relativistic} treatment \cite{sm72}
one uses the original single nucleon amplitude
given in terms of the Dirac spinors \cite{mrr69,bli73}.
In the non-relativistic approximation,
one uses the non-relativistic (NR) reduction
(the Foldy-Wouthuysen transformation)
of the single-particle transition amplitude \cite{wal75}.

In the NR approximation the cross sections for
eqs. (\ref{eq:nuAX}), (\ref{eq:nubarAX})
can be factorized into the single-nucleon cross section
and the nuclear response function which accounts for
the nucleon distribution inside nucleus and
the Pauli blocking effect.
Bugaev \etal. \cite{bugetal79}
used the NR form truncated at first order in
$ |\mbox{\bm $q$} |/M$, where \mbox{\bm $q$}
is the three-momentum transfer.
and calculated the inclusive cross sections on
$^{12}\rm{C}$, $^{16}\rm{O}$, $^{37}\rm{Cl}$,
$^{71}\rm{Ga}$ and $^{81}\rm{Br}$,
for $E_{\nu} \lsim$ 300 MeV.
The nuclear wavefunctions were taken to be
shell model wavefunctions in suitably chosen
Woods-Saxon potential.  Both bound and unbound orbits
were allowed as the final-state orbits.
Gaisser and O'Connell \cite{go86}
used the Fermi gas wavefunctions and
calculated the inclusive cross sections for
$E_{\nu}=$ 50 MeV $\sim$ 2 GeV.
To the extent that the Fermi gas model parameters
($p_F(p)$, $p_F(n)$, and
the average nucleon separation energy $E_B$)
can characterize actual nuclei,
the results can be applied to any nuclear targets.
Above the pion production threshold,
$X$ in eqs. (\ref{eq:nuAX}), (\ref{eq:nubarAX})
should include meson production processes, but
they were not considered in \cite{go86}.
(For neutrino-nucleus reactions
leading to meson productions,
see \cite{gnp86,anp74}.)
According to Gaisser and O'Connell \cite{go86},
(i) the relativistic and NR treatments give
{\it virtually identical} results,
(ii) in the energy region
where their and Bugaev \etal.'s  calculations overlap,
the results are {\it similar},
and (ii) the results are {\it accurate} to energies
as low as 50 MeV for electrons,
and to threshold for muons.
Thus, although Gaisser and O'Connell's calculation
\cite{go86} has been playing an extremely important
role in analyzing the $\nu_{\mu}/\nu_e$ ratio problem
\cite{hiretal88a,bioetal88},
the measure of possible errors in the results is not readily
available from \cite{go86}.

Kuramoto, Fukugita, Kohyama and Kubodera \cite{kfkk90}
made a detailed study of the
$\nu (\bar{\nu})$-$^{16}$O inclusive reactions
for $E_{\nu} \lsim$ 300 MeV,
using the relativistic Fermi gas (RFG) model
of Smith and Moniz \cite{sm72}
as well as an improved non-relativistic Fermi gas (NRFG) model
that incorporates the Foldy-Wouthuysen transformation
up to third order in $|\mbox{\bm $q$} |/M$.
One of the motivations for Kuramoto \etal. to go
beyond the lowest-order approximation of \cite{go86}
is that in some kinematical regions
the differential cross sections given in \cite{go86}
become negative, a rather disturbing aspect.
Kuramoto \etal.'s NRFG treatment is found to be free
from this problem and in reasonable agreement with
the RFG model results.
The NRFG cross sections systematically tend
to be somewhat smaller than
the RFG results and the difference,
which grows with $E_{\nu}$,
reaches $\sim$10 \% at $E_{\nu}$ = 300 MeV.
The agreement between the results of Kuramoto \etal.
and those of Bugaev \etal. \cite{bugetal79}
is also at the 10 \% level.

For lower incident energies,
there were attempts to add up explicitly
the contributions of individual final nuclear states.
Langworthy, Lamers and \"Uberall (LLU) \cite{llu77}
considered eighteen final states for
$\nu_{\ell}+ ^{16}{\rm{O}} \ra \ell^- +X({\rm{anything}})$,
and used the generalized Helm model
to parametrize individual transition strengths,
which were determined semi-empirically
with the use of the electron scattering data.
Haxton \cite{hax87} carried out
a summation of all possible final
$^{16}$F states contained
in a given truncated shell-model space;
all configurations up to $2\hbar \omega$ excitations
were included for positive-parity final states
and those up to $1 \hbar \omega$
excitations for negative-parity final states.
As in the empirical effective operator method (cf. chapter 4)
the shell-model transition rates were rescaled
by adjustable multiplicative factors
(up to $\sim 70 \%$ correction)
so that the empirical $B$(GT) values and
the electron scattering data were reproduced.
As discussed in Kuramoto \etal. \cite{kfkk90},
$\sigma_{\rm{Haxton}}$ agrees very well with
$\sigma_{\rm{RFG}}$ for $\Enu \lsim 60 \, \rm{MeV}$
but starts to deviate appreciably beyond
$\Enu \approx 60 \, \rm{MeV}$;
$\sigma_{\rm{LLU}}$ exhibits a similar tendency,
but its deviation from
$\sigma_{\rm{RFG}}$ beyond $\Enu \approx$ 60 MeV
is more pronounced.
Attributing these features to the fact
that the LLU \cite{llu77}
took into account only eighteen final nuclear states
whereas Haxton included much more,
Kuramoto \etal. argue that the difference between
$\sigma_{\rm{Haxton}}$ and
$\sigma_{\rm{RFG}}$ for $\Enu \gsim 60$ MeV
is due to the limited size of the shell-model space used by Haxton.
A later work of Haxton and Johnson \cite{hj90} drastically
extended the shell-model space
and calculated the $\beta$-decay
and $\mu$-capture.
It would be interesting to see what
cross section this extended calculation
will give for the inclusive reactions, eqs. (\ref{eq:nuAX}),
(\ref{eq:nubarAX}).

Although it is gratifying that the RGF results smoothly
join those of the explicit enumeration method
around $\Enu \approx$ 60 MeV,
all these methods share
the basic assumption that the quasi-free scattering
picture is valid.  The recent data
on the transverse and longitudinal nuclear responses,
$R_T$ and $R_L$,
in quasi-elastic ($e,e^{\prime}$) reactions indicate
significant deviations from the Fermi-gas prediction
even in the quasi-elastic peak region;
the discrepancy can be as large as
$\sim 20 \, \%$ for $^{12}$C
and
$\sim 50 \%$ for $^{40}$Ca.
Thus, the original remarkable success
\cite{mon69,whietal74,monetal71}
of the Fermi-gas model in predicting the
total (transverse +longitudinal)
response needs to be taken with caution.
There have been many attempts to explain
$R_T$ and $R_L$,
but a consistent picture is yet to be found \cite{cpv89}.
The fact that we are dealing here
with the total nuclear response is a favorable factor,
but the fact that we are calculating
the angle-integrated cross sections
that include the non-quasi-elastic region
may spoil to some extent the quasi-free scattering picture.
Taking the mass-number dependence of the above-mentioned
deviation from the Fermi gas model,
Kuramoto \etal. assigned $\sim 30 \%$ errors
to their estimation of the total cross sections for
$\nu_{\ell}+ ^{16}{\rm{O}} \ra \ell^- +X({\rm{anything}}) $
and
$\bar \nu_{\ell}+ ^{16}{\rm{O}} \ra \ell^+ +X({\rm{anything}}) $.

Singh and Oset \cite{so93} have recently
made an extremely detailed study
for $^{12}\rm{C}$ and$^{16}\rm{O}$,
using another version of the Fermi gas model,
the local Fermi gas model,
which was tested in several other processes.
Furthermore, these authors considered
the nuclear-medium quenching of the
weak-interaction effective coupling constants
and the final-state Coulomb distortion effect.
Comparison of Singh and Oset's results
with those of Kuramoto \etal. also indicates
that the Fermi gas model is probably reliable
up to the $\sim$ 30 \% level.

In actual experiments using a water \v{C}erenkov detector,
$\nu_{\ell}$ and $\bar \nu_{\ell}$
can also scatter from electrons.
It is however easy to separate
these events from
$\nu$ ($\bar \nu$) scattering off nuclei,
since the former has a very sharp forward peak
within $\theta \approx (m_e/E_{\nu})^{1/2}$.
Also, if the incident flux consists of a roughly equal mixture of
$\nu_e$ and $\bar \nu_e$,
the event rates of the reactions off oxygen nuclei
will compete with that of
$\bar \nu_e + p \ra e^+ + n$,
and it is important to separate one from the other.
This can also be done using the angular distribution
\cite{kfkk90}.

The significance of the improved estimates of
the cross sections for
$^{16}{\rm{O}}(\nu_{\ell},\ell^-)X({\rm{anything}})$
and
$^{16}{\rm{O}}(\bar \nu_{\ell},\ell^+)X({\rm{anything}})$
for the solar-flare neutrino problem will be discussed
at the end of the next chapter.
As far as the atmospheric neutrino problem is concerned,
the question is to what extent
the ambiguity in the Fermi gas model used in
interpreting the data would affect the conclusion.
Although one generally expects
that a substantial part of the ambiguity
will disappear upon taking the $\nu_{\mu}/\nu_e$ ratio
\cite{man93},
a more quantitative consideration seems warranted.

\newpage

\noin
{\bf Chapter 7 \hspace{1cm}
Semi-inclusive neutrino-nucleus reactions}

\vspace{0.4cm}
\noin
{\bf 7.1.}

When high-energy neutrinos hit a nuclear target
used in a radiochemical experiment,
many nuclear levels can of course be excited
but the experiment by design registers
only the reactions leading to particle-stable states.
If the number of final bound states is small enough
we can hope to calculate the contributions
of the individual levels and sum them up.
(This might  be the case for very light nuclei.)
As $E_{\nu}$ gets higher, however,
the possible final bound states can be too numerous
for the explicit enumeration of the individual levels
to be practical.
We call this type of process the semi-inclusive reaction.
Semi-inclusive reactions
cannot be treated
in the ordinary inclusive-reaction formalisms
since particle-unbound states
must be excluded from the final states.
A practically interesting example of
semi-inclusive reactions was encountered when
Davis \cite{dav86} considered the reaction between
the solar-flare neutrinos and the $\Cl37$ target:
\beq
\nu_{\ell}\,+\, ^{37}\rm{Cl}
\ra \ell^- \, + \, ^{37}\rm{A} \; \;\;
(\ell =e  \; \; \rm{or} \; \; \mu), \lab{eq:NuelCl37}
\eeq
where the final $^{37}\rm{A}$ stands for all
particle-stable states.
Davis's original argument used a rather
crude estimate of this semi-inclusive cross section,
which motivated Kuramoto \etal. \cite{fkkk89,kfkk90}
to make a more quantitative estimation.
Although Kuramoto \etal.'s treatment
is highly specific to the A=37 system,
we briefly describe that work here
as a possibly useful example,
since semi-inclusive reactions are likely
to be relevant to other radiochemical experiments as well,

The general expression
for cross section $\sigma$ for the reaction
$\nu_{\ell}+ |i> \ra \ell^- + |f>$ is
\beq
\sigma = (2\pi)^4 \sum_{f} \int d^3p_{\ell}
         \delta (E_{\ell}+E_f-E_{\nu}-E_i) \,
         |\!<\ell (p_{\ell});f|H_{\rm{eff}}|\nu(p_{\nu});i>\!|^2.
\lab{eq:}
\eeq
The final nuclear states $|f>$ should lie below
the neutron threshold energy, $E_x=8.8$ MeV.
In impulse approximation to be used here,
$H_{\rm{eff}}$ is the sum of the single-particle
weak-interaction operator.
The precision of the calculation described here
does not warrant taking account of
the exchange current effects
and other fine details.
In this approximation
\beq
\sigma = \frac{G^2}{\pi}\cos^2\theta_c
       \sum_{f} p_{\ell}E_{\ell}F(Z,E_{\ell}) \frac{1}{2}
       \int_{-1}^{1} d(\cos\theta) \, M_{\beta},
\eeq
where $E_{\ell} = \Enu -\Delta E_{fi}$
($\Delta E_{fi} \equiv E_f -E_i$) is the outgoing lepton energy.
The squared nuclear transition matrix elements $M_{\beta}$
is given by:
\beq
M_{\beta} = M_F |<f| \tilde {1} |i>|^2 +
            M_{G0} \,\frac {1}{3}\, |<f| \tilde {\sigma} |i>|^2 +
            M_{G2}\Lambda
\eeq
where
\ben
|<f| \tilde {1} |i>|^2 &\equiv& \frac{1}{2J_i+1} \sum_{M_f,M_i}
    |<f;J_f,M_f|\sum_{k=1}^{A}t_+(k)
  e^{i \mbox{\bm $q \cdot r$} _k}  |i;J_i,M_i>|^2  \\
 &= & \frac{4\pi}{2J_i+1} \sum_{\ell}
|<J_f||\sum_{k}t_+(k)j_{\ell}(qr_k)
Y_{\ell}(\hat {\mbox{\bm $r$}} _k) || J_i>|^2,
\lab{eq:tilde1}
\een
\ben
|<f| \tilde {\sigma} |i>|^2 &\equiv&
\frac{1}{2J_i+1} \sum_{M_f,M_i}
|<f;J_f,M_f|
\sum_{k=1}^{A}t_+(k) \mbox{\bm $\sigma$} (k)
e^{i \mbox{\bm $q \cdot r$} _k} |i;J_i,M_i>|^2  \\
 &= & \frac{4\pi}{2J_i+1} \sum_{\ell,K}
|<J_f||\sum_{k}t_+(k)j_{\ell}(qr_k)
[Y_{\ell}(\hat {\mbox{\bm $r$}} _k)
\times \mbox{\bm $\sigma$}(k)]^{(K)} || J_i>|^2,
\lab{eq:tildesigma}
\een
and $\Lambda$ represents the interference between
the $<\!f| \tilde {1} |i\!>$ and
$<\!f| \tilde {\sigma} |i\!>$ terms.
The coefficients, $M_F$, $M_{G0}$ and $M_{G2}$,
contain the nucleon form factors
as well as the kinematical quantities \cite{kfkk90}.
Now, of the transition operators
appearing in eqs. (\ref{eq:tilde1}), (\ref{eq:tildesigma}),
the operators that satisfy the selection rule:
``$\Delta J=0 \; \rm{or} \; 1, \; \Delta \pi \; =\, \rm{no}$"
are classified as allowed-type operators.
All the other operators are grouped together as forbidden-type.
(This classification slightly differs from what is customarily
used in $\beta$ decay theory.)
In the present case with
$|i\!>\,= |^{37}\rm{Cl}(\rm{gnd});\frac {3}{2} ^+\!>$,
the spin-parity of particle-bound states $|f\!>$
that can be reached
via allowed-type operators is $\frac {1}{2}^+$,
$\frac {3}{2}^+$, or $\frac {5}{2}^+$.
The assembly of these states will be referred to as group A.
Meanwhile, particle-bound states that can be fed only
through the forbidden-type operators constitute group B,
which contains negative-parity states as well as higher-spin
positive-parity states.

For those levels of group A
that predominantly belong to
sd-shell configurations,
reasonably realistic wave functions are available
from the work of Wildenthal \cite{wil84}.
These wave functions are obtained
by diagonalizing full sd-shell configurations and,
with the use of the empirical effective
operator method,
the Wildenthal wave functions generally
lead to a satisfactory agreement
with the observed $B$(GT)'s \cite{wil84,bw85}.
The result of this detailed rescaling
of the strengths can be simulated reasonably well
by replacing the free-nucleon value $g_A=-1.262$
with an effective value
$g_A^{\rm{eff}} \approx -(0.7 \sim 1.1)$
\cite{ah87}.
At the time of Kuramoto \etal.'s work,
both the $B$(GT) values determined
from the $^{37}{\rm{Cl}}(p,n)^{37}{\rm{Ar}}$ reaction and
those deduced from
the $^{37}{\rm{Ca}} \ra \,^{37}{\rm{K}} + \beta^+$ decay
had some irregularities
either among themselves or as compared
with the shell-model predictions.
Kuramoto \etal. \cite{kfkk90}
used as the reference GT strengths
the average of the $B$(GT) values determined
from these two methods,
except that,
for the $E_x \geq 6.5 \, \rm{MeV}$ region,
they adopted the (p,n) data in view of
the Adelberger and Haxton's argument \cite{ah87}
and, in the $E_x \approx 5 \, \rm{MeV}$ region,
they used the $B$(GT)'s obtained
from the $^{37}\rm{Ca}$ $\beta$-decay
with their strengths slightly reduced according to
\cite{ah87}.
Comparison of this reference $B$(GT)
with the shell model prediction indicates that
the shell model value
(with $g_A^{\rm{eff}} \approx -1.00$) slightly overestimates
$B$(GT) for $E_x <4 \, \rm{MeV}$
(by a factor of 1.38 on the average)
and slightly underestimate it
for $E_x > 4 \,\rm{MeV}$ (by a factor of 0.78).
In order to reproduce $B$(GT) for $E_x <4 \, \rm{MeV}$,
one needs to use $g_A^{\rm{eff}}=-0.85$.
Kuramoto \etal. assumed that this coupling constant
could be used universally for the whole range of $E_x$.
This implies that the average shell-model value of $B$(GT)
for $E_x > 4 \, \rm{MeV}$ is smaller
than the experimental value
by a factor of $0.78 \times 0.72 = 0.56$.
This deficit was interpreted as representing the contribution
of configurations outside the sd-shell.
Using this information on $B$(GT),
Kuramoto \etal. calculated
$\sigma_A$ as
\beq
\sigma_A=\sigma_{analog}+\sum_{E_x \leq 4\rm{MeV}}
                      \;  +\; \alpha \sum_{4< E_x <8.8 \rm{MeV}},
\eeq
where all the axial-vector contributions were calculated
for the Wildenthal wave functions using $g_A^{\rm{eff}}=-0.85$,
and $\alpha \equiv \frac{1}{0.56} =1.77$
was introduced to ``recover" missing contributions
of the outside-sd-shell contributions.

The levels belonging to group B may be
divided into two subgroups:
group $B_{sd}$ consisting of
positive-parity states with $J_f \geq \frac {7}{2}$
which belong to the sd-shell,
and group B' that comprises all others.
The contribution of group $B_{sd}$, denoted by
$\sigma_B^{sd}$, is estimated
using the Wildenthal wave functions and
$g_A^{\rm{eff}}=-0.85$.
The contribution of group B', denoted by $\sigma_{B'}$,
is assessed by using
$\sigma_{B'} = \sigma^{\rm{forbid}}N_{B'}$,
where $\sigma^{\rm{forbid}}$ is an estimate of
the typical cross section to
excite a level of group B', and $N_{B'}$
the estimated number of levels belonging to group B'.
The $\sigma^{\rm{forbid}}$ was estimated by taking the average of
$\sigma_B^{sd}$ and $(\sigma_{A'}^{sd})_{\rm{forbid}}$,
where $(\sigma_{A'}^{sd})_{\rm{forbid}}$
is defined through the decomposition
$\sigma_{A'}^{sd} =(\sigma_{A'}^{sd})_{\rm{allowed}}
                    + (\sigma_{A'}^{sd})_{\rm{forbid}}$,
with
$(\sigma_{A'}^{sd})_{\rm{allowed}}$
[$(\sigma_{A'}^{sd})_{\rm{forbid}}$]
being the contribution of
the allowed-type [forbidden-type] operators.
The assessment of $N_{B'}$ is a delicate job,
but referring to the available information \cite{ls78}
on the level structure in $\A37$,
Kuramoto \etal. estimated $N_{B'} \approx 28 \sim 34$.

The total contribution is obtained as
\beq
\sigma = \sigma_A + \sigma_B
= \sigma_A\, +\, \sigma_B^{sd} \, + \, \sigma_{B'}.
\eeq
Using this expression, Kuramoto \etal. \cite{kfkk90}
calculated the cross section for the semi-inclusive process
eq. (\ref{eq:NuelCl37}).
The cross section rises up to 150 $\sim$ 200 MeV
and then levels off.
This is a generally expected behavior for
a semi-inclusive reaction feeding bound states,
since the growth of the phase space factor
is counterbalanced by a suppression
due to the nuclear-size effect.
Kuramoto \etal. estimated the error of their
results to be 20 \% at 50 MeV,
40-50 \% at 100 MeV, and 50-60 \% or more at 150 MeV.
The cross section for $E_{\nu}> 150 \,\rm{MeV}$
may be subject to a factor 2 uncertainty.
For the use in calibration experiments with a muon beam,
the cross section averaged over the neutrino-spectrum
from the stopped muon decay is relevant.
Kuramoto \etal. \cite{kfkk90} give
$\sigma(\Cl37)_{\mu^+-{\rm{decay}}} =
(9.4\pm1.4) \times 10^{-41}\,{\rm{cm}}^2$.
This value should be compared with
$9.90 \times 10^{-41}\,{\rm{cm}}^2$ given
in \cite{dh77}, and with
$7.6 \times 10^{-41}\,{\rm{cm}}^2$ \cite{bu88}.

Although the method used by Kuramoto \etal.
is far from systematic,
it is at present difficult to do better.
It is hoped that the progress of
main-stream shell model calculations
will eventually reach the stage wherein
the adopted model space is sufficiently
large to accommodate all important final states
of semi-inclusive neutrino-nucleus reactions,
at least up to $sd$-shell nuclei.
These shell-model model wavefunctions plus
a judicious use of the empirical effective operator method
(chapter 5)
will probably allow us to treat the semi-inclusive
reactions in a more systematic way.
For nuclear targets beyond the $sd$-shell,
the possibility of such a systematic treatment
seems rather remote.

\vspace{0.4cm}
\noin
{\bf 7.2.}\hspace{0.3cm}
Consequences for the solar-flare problem

Kuramoto \etal. \cite{kfkk90}
used their estimate of the semi-inclusive reaction
eq. (\ref{eq:NuelCl37}) to deduce
the solar-flare neutrino flux
corresponding to the fifty excess $\A37$ events
in the 615 ton $\rm{C}_2\rm{Cl}_4$ Homestake detector;
this excess was assumed
by Davis \cite{dav86} to be of the solar-flare origin.
For the flare-neutrino spectrum,
a soft neutrino spectrum of the form
$\phi(E_{\nu}) = \phi_0 \, \exp(-E_{\nu}/E_0)$,
with $E_0=50$ MeV, as well as a monoenergetic
form with various plausible energies was used.
Based on the solar-flare neutrino flux deduced therefrom
and their estimate (chapter 5)
of the cross section for
$^{16}\rm{O}(\nu_{\ell},\ell^-)X(\rm{anything})$,
Kuramoto \etal. calculated
the number of neutrino events to be expected
in the Kamiokande detector during the solar
flare periods and concluded that
the observed counting rate of the neutrino events
\cite{hiretal88a}
fell far below the prediction.
This corroborates the earlier conclusion
of the Kamiokande group \cite{hiretal88b},
that the time-integrated ``solar-flare $\nu_e$" flux
is less than $3.7 \times 10^7 \rm{cm}^{-2}$ per flare for
$E_{\nu}=100$ MeV, which is 2000 times smaller
than the value required to account for
the excess ``signal" observed in the $\Cl37$ experiment.
Thus, the role of the solar-flare neutrinos in giving rise to
the time structure in the Homestake solar-neutrino
data must be much smaller than described in \cite{dav86}.
For further detailed studies on the statistical significance
of the time structure in the solar-neutrino data,
see \cite{bp91,kra87,bieetal90}.
The untenability of Davis's original suggestion
does not diminish the significance of
using water \v{C}erenkov counters,
like the Kamiokande or IMB,
to monitor intermediate energy
neutrinos associated with the solar activity.
This will allow us to explore the active Sun
with a new, unconventional means,
and also to have a better understanding
of the particle acceleration
mechanism around the solar sphere.
It is to be emphasized
that the sensitivity of the
Kamiokande and IMB water detectors reaches
the flux level of an optimistic estimate for a large flare
or for more constant activity
during the solar maximum period \cite{bfp87,kov81}.

\newpage

\noin
{\bf Chapter 8 \hspace{1cm} Additional remarks}

In the foregoing chapters
we have discussed various nuclear physics aspects
that go into the estimation of neutrino-nucleus reactions
of astrophysical interest.
Neutrino-nucleus reactions can also be useful
in studying the structure and symmetry properties
of the fundamental coupling of the weak current with matter.
In some of these experiments nuclei serve
just as an assembly of $A$ nucleons,
whereas in many others
one takes advantage of a wide choice of selection rules
pertaining to specific nuclear targets.

An example of the former is the recent determination
of a stringent upper limit to the second-class current
by Ahrens \etal. \cite{ahr88}.
These authors measured quasielastic reaction
$\bar{\nu}_{\mu}p \ra \mu^{+}n$ occurring in
the scintillation target detectors for $q^2$ up to
1.0 (GeV/$c)^2$.
For the second-class axial-vector current strength
defined by
\beq
<\! n(p_f)| A_{\mu}^{\rm{SCC}}|p(p_i) \!> =
{\rm{i}} \frac{f_T}{M}\,\bar{u}(p_f)\gamma_5
\sigma_{\mu\nu}q^{\mu}u(p_i)
\eeq
with $q \equiv p_f-p_i$,
and for the dipole parametrization of
the $q^2$-dependence
$f_T(q^2) = \eta f_A(0)/(1+q^2/M_T^2)^2$,
Ahrens \etal. deduced $|\eta| \leq$ 0.8, 0.23, 0.12
for $M_T$(GeV/c$^2)=$ 0.5, 0.9, 1.5, respectively.
These upper limits are substantially lower than
the value obtained from analyzing the totality of
the existing $\beta$ decay data. \cite{ok80,gre85}.
To the extent that the incoherent sum of the individual
quasielastic scattering off a nucleon provides an accurate
description of the process, no nuclear dynamics is involved here.
Although this picture is likely to be valid,
it is probably worthwhile to check its validity
in a quantitative manner in view of
the breakdown (up to 30\%) of the quasi-free scattering
assumption in the longitudinal and transverse
response functions in $(e,e^{\prime})$ processes.

There have been numerous attempts at exploiting
various selection rules available
from specific nuclear quantum numbers
to study unconventional pieces of the weak current.
Although it is beyond the scope of this review
to discuss them in detail,
we mention here as an example
a possible consequence of the strangeness content
of the nucleon for neutrino-nucleus reactions.
Although this effect will {\it not} change drastically
the interpretation of astrophysical neutrino-nucleus reactions,
it has its own fundamental importance.

The strange quark matrix elements of the nucleon
are one of the central issues in contemporary
particle-nuclear physics.
The measurement of polarized
deep-inelastic muon scattering on the proton \cite{ash88},
combined with an analysis of
hyperon $\beta$--decay \cite{bou83},
suggests that there is a large contribution
of the strange sea quarks ($\bar{s}s$) to the proton spin.
The pion-nucleon sigma term $\Sigma_{\pi N}$
also indicates a significant
amplitude of $<\!N|\bar{s}s|N\!>$,
although the latest estimate \cite{gls91}
of $\Sigma_{\pi N}$ favors somewhat smaller $\bar{s}s$
content than in \cite{dn86}.
Beise and McKeown \cite{bei91} have discussed
parity violation measurements of
polarized electron--proton scattering
to study strange quark matrix  elements
of the vector current.
If such a large $\bar{s}s$ component exists in the nucleon,
one should expect a sizable contribution also in the
neutrino--nucleus reactions.
In fact, neutrino neutral--current reactions
offer the best means
to study the strange quark matrix element
of the axial current,
which complement the electron-scattering experiments
that are sensitive only to the vector current.

In the presence of the strange quarks,
the hadronic neutral current $J_{\mu}^3$
appearing in eq. (\ref{eq:Hnc})
would become $J_{\mu}^{\rm{NC}}$, where
\beq
J_{\mu}^{\rm{NC}} = \, J_{\mu}^{T=1} \, + \, J_{\mu}^{T=0} \, ,
\eeq
with the isovector and isoscalar pieces defined by
\ben
J_{\mu}^{T=1} \, &= \, ( 1 - 2 \sin^{2} \theta_{W} )
\, V_{\mu}^3 \, + \, A_{\mu}^3  \\
J_{\mu}^{T=0} \, &= \, - 2 \sin^{2} \theta_{W} \,
V_{\mu}^0 \,
+ \, V_{\mu}^s \, + \, A_{\mu}^s\>.
\een
The standard forms for the effective isovector currents
in the nucleon sector are
\ben
V_{\mu}^{(3)} \, &=& \, \bar{u}_{N'} \left( F_{1}^{V} \gamma_{\mu} \,
 + \, F_{2}^{V}
\displaystyle{\frac{i \sigma_{\mu \nu}q^{\nu}}
{2 m_{N}}} \right) \displaystyle{\frac{\tau_{3}}{2}} u_{N}  \\
A_{\mu}^{(3)} \,&=& \,
g_{A} \bar{u}_{N'} \gamma_{\mu} \gamma_{5}
\displaystyle{\frac{\tau_{3}}{2}} u_{N}     \, ,
\een
where $F_{1}^{V}$ and $F_{2}^{V}$ are isovector form factors,
and $g_{A}(0) = -1.26$.
Similarly, the isoscalar and the strange quark currents
\cite{bei91} are expressed as
\ben
V_{\mu}^{(0)} \, &=&
 \, \bar{u}_{N'} \left( F_{1}^{0} \gamma_{\mu} \,
 + \, F_{2}^{0} \displaystyle{\frac{i \sigma_{\mu \nu}q^{\nu}}
{2 m_{N}}} \right)    \displaystyle{\frac{1}{2}} u_{N}    \\
V_{\mu}^{(s)} \, &=& \,
- \displaystyle{\frac{1}{2}} \bar{u}_{N'}
\left( F_{1}^{s} \gamma_{\mu} \, + \, F_{2}^{s}
\displaystyle{\frac{i\sigma_{\mu \nu}q^{\nu}}{2 m_{N}}} \right)
\displaystyle{\frac{1}{2}}   u_{N}     \\
A_{\mu}^{(s)} \, &=& \,
\displaystyle{\frac{1}{2}} G_{1}^{s} \bar{u}_{N'}
\gamma_{\mu} \gamma_{5} \displaystyle{\frac{1}{2}} u_{N}   \, ,
\een
where $F_{1}^{0}$ and $F_{2}^{0}$ are isoscalar form factors, and
$F_{1}^{s}$,
$F_{2}^{s}$ and $G_{1}^{s}$ denote the strange quark form factors
of the nucleon for the vector current
and for the axial vector current, respectively.
Attempts to extract the strange quark axial form factor $G_{1}^{s}(0)$
have been made using both EMC \cite{ash88,bou83} and $\nu$--$p$
scattering
data \cite{ahr87,kap88}. The extracted form factors are
\beq
 G_{1}^{s}(0) \,  = \, \cases{
      -0.38 \pm 0.11 &\ \ \ \ \ \ \ \ from EMC    \cr
      -0.30 \pm 0.16 &\ \ \ \ \ \ \ \ from $\nu$--$p$.  \cr  }
\eeq
However, it should be emphasized that the quoted numbers are not
yet well
established, because of several assumptions made in the analyses
\cite{kap88},
so that the quoted errors are the experimental ones {\it only}.
An interesting question is to what extent
neutrino-nucleus reactions can sharpen our conclusion
on the strange-quark neutral current form factors.
Henley \etal. \cite{hen91} have investigated the asymmetry
between the elastic cross--sections for
$\nu + d \ra \nu + d $ and
$\bar{\nu} + d \ra \bar{\nu} + d$.
In impulse approximation,
the difference in cross sections is given by
\ben
N \, \equiv \,
\displaystyle{\frac{1}{2}}
\left\{ \displaystyle{\frac{d \sigma}{d q^{2}}}
(\nu d \ra \nu d) \, - \,
\displaystyle{\frac{d \sigma}{d q^{2}}}
(\bar{\nu} d \ra \bar{\nu} d) \right\} \\
= \, \displaystyle{\frac{4}{3}}
\displaystyle{\frac{G_{F}^{2} m_{N}}{\pi E}}
\eta (1 - \zeta) G_{1}^{s}
\left( F_{1}^{0} + F_{2}^{0} \right) F_{D} \, ,
\een
where $\eta = q^{2}/4 m_{N_{\nu}}^2$,
$\zeta = q^{2}/8m_{N}E_{\nu}$,
with $q$ the momentum transfer.
$E_{\nu}$ is the neutrino (anti-neutrino) energy,
and $F_{D}$ is the deuteron form factor.
According to Henley \etal.'s calculation
the asymmetry $R$ defined by
\beq
R \, \equiv \,
\displaystyle{ \frac{(d \sigma /d q^{2}) (\nu d \ra \nu d)
\, - \, (d \sigma /d q^{2})(\bar{\nu} d \ra \bar{\nu} d)}
{(d \sigma /d q^{2})(\nu d \ra \nu d) \, + \,
(d \sigma /d q^{2}) (\bar{\nu} d \ra \bar{\nu} d)}}
\eeq
could be as large as 15 \% for
large $q^{2}$ ($\geq$ 0.5 GeV$^{2}/c^{2}$).
Suzuki, Kohyama and Yazaki\cite{sky90}
made a detailed study of low-energy
$\nu$--$^{12}$C and $\nu$--$^{10}$B reactions
to extract information on $G_1^s$.
Bernab\'{e}u {\it et al.} \cite{ber92} studied
the effect of the strange quark matrix element
for the neutral-current reaction
\beq
\nu \,+ \,^{11}{\rm B} \ra \nu \,+ \,^{11}{\rm B}^{*} \, .
\eeq
This reaction is particularly interesting,
because it is directly related
to BOREX or BOREXINO experiment \cite{rpb86}.
It is to be noted that, unlike the $\nu d$ reactions,
the isovector and isoscalar currents contribute
coherently to the cross section here.
A similar situation should exist for the
$\nu^{13}{\rm{C}}$ neutral-current reaction
discussed in chapter 4.

The above considerations were all based on the impulse approximation.
Very recently, Hjorth-Jensen, Kirchbach, Riska and Tsushima
\cite{hjoetal93} have pointed out that the exchange currents
can make the effective isoscalar axial current density of a nucleon
in a nucleus considerably different
from the corresponding free nucleon value.
The isoscalar axial charge coupling (time component) can be
enhanced by 40-50 \% due to the short-range component
of the nucleon-nucleon interaction,
whereas the isoscalar axial current coupling (space component)
is reduced by about 10\% by the exchange current involving
the $\pi a_0(980)$ exchange mechanism.
It seems interesting to examine
the consequences of these effective coupling constants
for neutrino-nucleus scatterings.

\vspace{0.5cm}

To summarize, we wish to reemphasize that
neutrino-nucleus reactions play an important role
in the study of the astrophysical neutrinos.
The diversity of possible types of reactions
presents a great challenge to nuclear physicists.
It is hoped that the present review
have succeeded in calling
our fellow nuclear physicists' attention to
some of the major problems facing us in this field.

\vspace{0.8cm}
\noin
{\bf Acknowledgement}

We wish to express our sincere thanks
to Dr. Y. Kohyama for many valuable consultations
on the topics discussed in this review article.
At various stages of our work on
neutrino-nucleus reactions,
we benefited greatly from collaborations
and/or discussions with Prof. M. Fukugita,
Prof. J. N. Bahcall, Prof. W. C. Haxton
and Prof. M. Koshiba,
to whom we would like to express our gratitude.
We are also indebted to Prof. C. Goodman for useful
comments on an early version of the manuscript.

\newpage

\setlength{\textheight}{23.0cm}
\setlength{\textwidth}{17.0cm}
\setlength{\topmargin}{-2.0cm}

\begin{table}[hbtp]

\noindent
Table 1 \hspace{1cm}
Neutrino-deuteron cross sections in units of cm$^2$
\vspace{0.8cm}

\begin{tabular}{|c|c|c|c|c|} \hline

$E_{\nu}$ (MeV) & $\nu d \rightarrow \nu'p n$  & $\bar{\nu} d
\rightarrow
\bar{\nu}' p n$ & $\nu d \rightarrow e^{-} p p$ &
$\bar{\nu} d \rightarrow e^{+} n n$ \\   \hline
          2.0  &  0.000E+00  &  0.000E+00  &  3.854E-45  &  0.000E+00
\\
          2.2  &  0.000E+00  &  0.000E+00  &  8.231E-45  &  0.000E+00
\\
          2.4  &  4.394E-47  &  4.362E-47  &  1.461E-44  &  0.000E+00
\\
          2.6  &  4.288E-46  &  4.253E-46  &  2.318E-44  &  0.000E+00
\\
          2.8  &  1.464E-45  &  1.451E-45  &  3.413E-44  &  0.000E+00
\\
          3.0  &  3.369E-45  &  3.334E-45  &  4.761E-44  &  0.000E+00
\\
          3.2  &  6.308E-45  &  6.236E-45  &  6.377E-44  &  0.000E+00
\\
          3.4  &  1.041E-44  &  1.028E-44  &  8.274E-44  &  0.000E+00
\\
          3.6  &  1.579E-44  &  1.557E-44  &  1.046E-43  &  0.000E+00
\\
          3.8  &  2.252E-44  &  2.219E-44  &  1.296E-43  &  0.000E+00
\\
          4.0  &  3.068E-44  &  3.021E-44  &  1.577E-43  &  0.000E+00
\\
          4.2  &  4.034E-44  &  3.967E-44  &  1.890E-43  &  1.065E-45
\\
          4.4  &  5.154E-44  &  5.064E-44  &  2.237E-43  &  4.397E-45
\\
          4.6  &  6.434E-44  &  6.314E-44  &  2.618E-43  &  9.832E-45
\\
          4.8  &  7.877E-44  &  7.722E-44  &  3.033E-43  &  1.748E-44
\\
          5.0  &  9.487E-44  &  9.290E-44  &  3.485E-43  &  2.747E-44
\\
          5.2  &  1.127E-43  &  1.102E-43  &  3.973E-43  &  3.993E-44
\\
          5.4  &  1.322E-43  &  1.292E-43  &  4.498E-43  &  5.498E-44
\\
          5.6  &  1.534E-43  &  1.498E-43  &  5.060E-43  &  7.270E-44
\\
          5.8  &  1.765E-43  &  1.721E-43  &  5.661E-43  &  9.319E-44
\\
          6.0  &  2.013E-43  &  1.961E-43  &  6.239E-43  &  1.165E-43
\\
          6.2  &  2.279E-43  &  2.218E-43  &  6.904E-43  &  1.427E-43
\\
          6.4  &  2.563E-43  &  2.491E-43  &  7.606E-43  &  1.719E-43
\\
          6.6  &  2.866E-43  &  2.782E-43  &  8.347E-43  &  2.040E-43
\\
          6.8  &  3.198E-43  &  3.102E-43  &  9.127E-43  &  2.392E-43
\\
          7.0  &  3.541E-43  &  3.430E-43  &  9.945E-43  &  2.774E-43
\\
          7.2  &  3.902E-43  &  3.776E-43  &  1.080E-42  &  3.187E-43
\\
          7.4  &  4.283E-43  &  4.140E-43  &  1.170E-42  &  3.631E-43
\\
          7.6  &  4.683E-43  &  4.522E-43  &  1.264E-42  &  4.106E-43
\\
          7.8  &  5.103E-43  &  4.921E-43  &  1.361E-42  &  4.612E-43
\\ \hline
\end{tabular}
\end{table}

\newpage

\begin{table}[hbtp]

\noindent
Table 1 (continued)\hspace{0.5cm}
Neutrino-deuteron cross sections in units of cm$^2$
\vspace{1cm}

\begin{tabular}{|c|c|c|c|c|} \hline
$E_{\nu}$ (MeV) & $\nu d \rightarrow \nu'p n$  & $\bar{\nu} d
\rightarrow
\bar{\nu}' p n$ & $\nu d \rightarrow e^{-} p p$ &
$\bar{\nu} d \rightarrow e^{+} n n$ \\   \hline
          8.0  &  5.542E-43  &  5.339E-43  &  1.463E-42  &  5.150E-43
\\
          8.2  &  6.000E-43  &  5.775E-43  &  1.569E-42  &  5.719E-43
\\
          8.4  &  6.479E-43  &  6.228E-43  &  1.679E-42  &  6.320E-43
\\
          8.6  &  6.977E-43  &  6.700E-43  &  1.793E-42  &  6.967E-43
\\
          8.8  &  7.495E-43  &  7.189E-43  &  1.911E-42  &  7.635E-43
\\
          9.0  &  8.033E-43  &  7.697E-43  &  2.033E-42  &  8.336E-43
\\
          9.2  &  8.591E-43  &  8.223E-43  &  2.160E-42  &  9.068E-43
\\
          9.4  &  9.169E-43  &  8.767E-43  &  2.290E-42  &  9.833E-43
\\
          9.6  &  9.767E-43  &  9.328E-43  &  2.425E-42  &  1.063E-42
\\
          9.8  &  1.039E-42  &  9.908E-43  &  2.564E-42  &  1.146E-42
\\
         10.0  &  1.103E-42  &  1.051E-42  &  2.708E-42  &  1.232E-42
\\
         10.2  &  1.168E-42  &  1.112E-42  &  2.855E-42  &  1.322E-42
\\
         10.4  &  1.236E-42  &  1.176E-42  &  3.007E-42  &  1.415E-42
\\
         10.6  &  1.306E-42  &  1.241E-42  &  3.163E-42  &  1.511E-42
\\
         10.8  &  1.378E-42  &  1.308E-42  &  3.323E-42  &  1.610E-42
\\
         11.0  &  1.453E-42  &  1.377E-42  &  3.488E-42  &  1.712E-42
\\
         11.2  &  1.529E-42  &  1.447E-42  &  3.657E-42  &  1.818E-42
\\
         11.4  &  1.607E-42  &  1.520E-42  &  3.831E-42  &  1.927E-42
\\
         11.6  &  1.688E-42  &  1.594E-42  &  4.008E-42  &  2.039E-42
\\
         11.8  &  1.770E-42  &  1.670E-42  &  4.191E-42  &  2.155E-42
\\
         12.0  &  1.854E-42  &  1.748E-42  &  4.377E-42  &  2.273E-42
\\
         12.5  &  2.075E-42  &  1.950E-42  &  4.863E-42  &  2.584E-42
\\
         13.0  &  2.308E-42  &  2.164E-42  &  5.378E-42  &  2.914E-42
\\
         13.5  &  2.555E-42  &  2.389E-42  &  5.920E-42  &  3.265E-42
\\
         14.0  &  2.815E-42  &  2.625E-42  &  6.491E-42  &  3.635E-42
\\
         14.5  &  3.088E-42  &  2.872E-42  &  7.090E-42  &  4.025E-42
\\
         15.0  &  3.375E-42  &  3.130E-42  &  7.719E-42  &  4.435E-42
\\
         15.5  &  3.675E-42  &  3.399E-42  &  8.376E-42  &  4.864E-42
\\
         16.0  &  3.989E-42  &  3.679E-42  &  9.063E-42  &  5.313E-42
\\
         16.5  &  4.316E-42  &  3.969E-42  &  9.780E-42  &  5.780E-42
\\
         17.0  &  4.656E-42  &  4.271E-42  &  1.053E-41  &  6.267E-42
\\
         17.5  &  5.010E-42  &  4.584E-42  &  1.130E-41  &  6.773E-42
\\
         18.0  &  5.378E-42  &  4.907E-42  &  1.211E-41  &  7.297E-42
\\
         18.5  &  5.760E-42  &  5.241E-42  &  1.295E-41  &  7.841E-42
\\
         19.0  &  6.155E-42  &  5.585E-42  &  1.381E-41  &  8.402E-42
\\
         19.5  &  6.564E-42  &  5.940E-42  &  1.471E-41  &  8.983E-42
\\ \hline
\end{tabular}
\end{table}

\newpage

\begin{table}[hbtp]

\noindent
Table 1 (continued)\hspace{0.5cm}
Neutrino-deuteron cross sections in units of cm$^2$
\vspace{1cm}

\begin{tabular}{|c|c|c|c|c|} \hline
$E_{\nu}$ (MeV) & $\nu d \rightarrow \nu'p n$  & $\bar{\nu} d
\rightarrow
\bar{\nu}' p n$ & $\nu d \rightarrow e^{-} p p$ &
$\bar{\nu} d \rightarrow e^{+} n n$ \\   \hline

         20.0  &  6.987E-42  &  6.306E-42  &  1.564E-41  &  9.581E-42
\\
         20.5  &  7.424E-42  &  6.682E-42  &  1.661E-41  &  1.020E-41
\\
         21.0  &  7.875E-42  &  7.069E-42  &  1.760E-41  &  1.083E-41
\\
         21.5  &  8.339E-42  &  7.466E-42  &  1.862E-41  &  1.149E-41
\\
         22.0  &  8.818E-42  &  7.873E-42  &  1.968E-41  &  1.216E-41
\\
         22.5  &  9.311E-42  &  8.291E-42  &  2.077E-41  &  1.285E-41
\\
         23.0  &  9.818E-42  &  8.719E-42  &  2.189E-41  &  1.355E-41
\\
         23.5  &  1.034E-41  &  9.157E-42  &  2.305E-41  &  1.427E-41
\\
         24.0  &  1.087E-41  &  9.606E-42  &  2.423E-41  &  1.501E-41
\\
         24.5  &  1.142E-41  &  1.006E-41  &  2.545E-41  &  1.577E-41
\\
           25  &  1.199E-41  &  1.053E-41  &  2.671E-41  &  1.655E-41
\\
           26  &  1.316E-41  &  1.150E-41  &  2.932E-41  &  1.815E-41
\\
           27  &  1.439E-41  &  1.251E-41  &  3.206E-41  &  1.981E-41
\\
           28  &  1.567E-41  &  1.355E-41  &  3.494E-41  &  2.155E-41
\\
           29  &  1.702E-41  &  1.464E-41  &  3.796E-41  &  2.335E-41
\\
           30  &  1.842E-41  &  1.576E-41  &  4.112E-41  &  2.521E-41
\\
           31  &  1.988E-41  &  1.692E-41  &  4.443E-41  &  2.713E-41
\\
           32  &  2.140E-41  &  1.812E-41  &  4.787E-41  &  2.912E-41
\\
           33  &  2.298E-41  &  1.936E-41  &  5.146E-41  &  3.117E-41
\\
           34  &  2.462E-41  &  2.063E-41  &  5.520E-41  &  3.328E-41
\\
           35  &  2.632E-41  &  2.194E-41  &  5.909E-41  &  3.545E-41
\\
           36  &  2.808E-41  &  2.329E-41  &  6.313E-41  &  3.768E-41
\\
           37  &  2.990E-41  &  2.467E-41  &  6.731E-41  &  3.996E-41
\\
           38  &  3.178E-41  &  2.609E-41  &  7.165E-41  &  4.231E-41
\\
           39  &  3.372E-41  &  2.754E-41  &  7.614E-41  &  4.470E-41
\\
           40  &  3.572E-41  &  2.902E-41  &  8.079E-41  &  4.716E-41
\\
           41  &  3.778E-41  &  3.055E-41  &  8.559E-41  &  4.966E-41
\\
           42  &  3.991E-41  &  3.210E-41  &  9.055E-41  &  5.222E-41
\\
           43  &  4.209E-41  &  3.368E-41  &  9.567E-41  &  5.484E-41
\\
           44  &  4.434E-41  &  3.530E-41  &  1.009E-40  &  5.750E-41
\\
           45  &  4.665E-41  &  3.695E-41  &  1.064E-40  &  6.021E-41
\\
           46  &  4.901E-41  &  3.864E-41  &  1.120E-40  &  6.298E-41
\\
           47  &  5.144E-41  &  4.035E-41  &  1.177E-40  &  6.579E-41
\\
           48  &  5.393E-41  &  4.210E-41  &  1.237E-40  &  6.865E-41
\\
           49  &  5.649E-41  &  4.387E-41  &  1.297E-40  &  7.156E-41
\\ \hline
\end{tabular}
\end{table}

\newpage

\begin{table}[hbtp]

\noindent
Table 1 (continued)\hspace{0.5cm}
Neutrino-deuteron cross sections in units of cm$^2$
\vspace{1cm}

\begin{tabular}{|c|c|c|c|c|} \hline
$E_{\nu}$ (MeV) & $\nu d \rightarrow \nu'p n$  & $\bar{\nu} d
\rightarrow
\bar{\nu}' p n$ & $\nu d \rightarrow e^{-} p p$ &
$\bar{\nu} d \rightarrow e^{+} n n$ \\   \hline

           50  &  5.910E-41  &  4.568E-41  &  1.360E-40  &  7.451E-41
\\
           51  &  6.177E-41  &  4.751E-41  &  1.424E-40  &  7.751E-41
\\
           52  &  6.451E-41  &  4.937E-41  &  1.490E-40  &  8.055E-41
\\
           53  &  6.731E-41  &  5.126E-41  &  1.557E-40  &  8.364E-41
\\
           54  &  7.017E-41  &  5.318E-41  &  1.626E-40  &  8.676E-41
\\
           55  &  7.309E-41  &  5.513E-41  &  1.697E-40  &  8.993E-41
\\
           60  &  8.860E-41  &  6.525E-41  &  2.076E-40  &  1.064E-40
\\
           65  &  1.056E-40  &  7.599E-41  &  2.498E-40  &  1.237E-40
\\
           70  &  1.241E-40  &  8.728E-41  &  2.961E-40  &  1.419E-40
\\
           75  &  1.440E-40  &  9.906E-41  &  3.467E-40  &  1.607E-40
\\
           80  &  1.654E-40  &  1.113E-40  &  4.014E-40  &  1.801E-40
\\
           85  &  1.881E-40  &  1.239E-40  &  4.603E-40  &  2.001E-40
\\
           90  &  2.121E-40  &  1.368E-40  &  5.232E-40  &  2.205E-40
\\
           95  &  2.373E-40  &  1.500E-40  &  5.900E-40  &  2.412E-40
\\
          100  &  2.637E-40  &  1.634E-40  &  6.606E-40  &  2.623E-40
\\
          110  &  3.199E-40  &  1.908E-40  &  8.127E-40  &  3.049E-40
\\
          120  &  3.801E-40  &  2.186E-40  &  9.784E-40  &  3.481E-40
\\
          130  &  4.437E-40  &  2.465E-40  &  1.156E-39  &  3.914E-40
\\
          140  &  5.103E-40  &  2.744E-40  &  1.345E-39  &  4.346E-40
\\
          150  &  5.793E-40  &  3.020E-40  &  1.543E-39  &  4.776E-40
\\
          160  &  6.501E-40  &  3.292E-40  &  1.750E-39  &  5.201E-40
\\
          170  &  7.224E-40  &  3.559E-40  &  1.963E-39  &  5.623E-40
\\ \hline
\end{tabular}
\end{table}

\end{document}